\documentclass[longauth,useAMS,usenatbib,onecolumn]{mn2e}
\usepackage{graphicx,bm,setspace,amssymb,color,amsmath,multirow}
\numberwithin{equation}{section}

\newcommand{\mv}[1]{{\bm #1}}
\newcommand{\ii}{\mathrm{i}}

\newcommand{\me}[1]{\exp \left[ {#1} \right]}
\newcommand{\fe}[1]{\me{\frac{\ii \upi}{\lambda h} {#1}}}
\newcommand{\ips}[1]{\left|\widetilde{\phi}\left({#1}\right)\right|^2}
\newcommand{\isf}[1]{\mathcal{D}\left({#1} \right)}
\newcommand{\lr}[1]{\left\langle #1 \right\rangle}
\newcommand{\beff}{B_{\rm eff}(d,\nu)}
\newcommand{\beam}{B(d,\nu,\mv{l})}
\newcommand{\seff}{S_{\rm eff}(d,\nu)}
\newcommand{\seffsq}{S^2_{\rm eff}(d,\nu)}
\newcommand{\smax}{S_{\rm max}(d,\nu)}
\newcommand{\sefd}{{\rm SEFD}(d,\nu)}
\newcommand{\bcdot}{\bm{\cdot}}

\title[Scintillation noise in widefield radio interferometry]
  {Scintillation noise in widefield radio interferometry} 

\author[Vedantham \& Koopmans]
{H.K.~Vedantham\thanks{E-mail: harish@astro.rug.nl} and L.V.E.~Koopmans\\
Kapteyn Astronomical Institute, University of Groningen, P.O. Box 800, 9700 AV Groningen, The Netherlands}

\begin{document}
%
\date{\today}
\pagerange{\pageref{firstpage}--\pageref{lastpage}} \pubyear{2014}
\def\LaTeX{L\kern-.36em\raise.3ex\hbox{a}\kern-.15em
    T\kern-.1667em\lower.7ex\hbox{E}\kern-.125emX}
\newtheorem{theorem}{Theorem}[section]
\label{firstpage}
\maketitle
%
%
%
%

\begin{abstract}
In this paper, we consider random phase fluctuations imposed during wave propagation through a turbulent plasma (e.g. ionosphere) as a source of additional noise in interferometric visibilities. We derive expressions for visibility variance for the wide field of view case (FOV$\sim10$~deg) by computing the statistics of Fresnel diffraction from a stochastic plasma, and provide an intuitive understanding. For typical ionospheric conditions (diffractive scale $\sim 5-20$~km at $150$~MHz), we show that the resulting ionospheric `scintillation noise' can be a dominant source of uncertainty at low frequencies ($\nu \lesssim 200$~MHz). Consequently, low frequency  widefield radio interferometers must take this source of uncertainty into account in their sensitivity analysis. We also discuss the spatial, temporal, and spectral coherence properties of scintillation noise that determine its magnitude in deep integrations, and influence prospects for its mitigation via calibration or filtering. 
\end{abstract}

%
%
%
%
\section{Introduction}
Low frequency radio astronomy ($50$~MHz $\lesssim \nu \lesssim 500$~MHz) is currently generating significant interest from across astronomical disciplines \citep{skabook}. In a build up to future telescopes such as the SKA\footnote{Square Kilometre Array: visit http://www.skatelescope.org for details} and HERA\footnote{Hydrogen Epoch of Reionization Array: visit http://reionization.org for details}, new pathfinder instruments such as LOFAR \citep{lofar}, MWA \citep{mwa}, GMRT \citep{gmrt}, and PAPER \citep{paper} are currently operational. Many of the science cases for these instruments demand unprecedented sensitivity levels. However, attaining the theoretical sensitivity limit dictated by thermal noise has been a perennial challenge at low frequencies ($\nu<200$~MHz). Low frequency radio waves are corrupted during their propagation through plasma in the interstellar and interplanetary media, and the Earth's ionosphere. Understanding the ensuing propagation effects is critical not only to mitigate the resulting systematic errors, but also to study the media themselves. These plasma are known to be turbulent in nature, and introduce a stochastic effect on radio wave propagation. In this paper, we treat this inherent randomness\footnote{We will call this phenomena as `visibility scintillation' after \citet{cronyn1972}. Manifestation of the same phenomenon in images will be called `speckle noise'.} as a source of uncertainty above and beyond the thermal noise. In doing so, we show that visibility scintillation due to ionospheric propagation can be a dominant source of uncertainty at low frequencies ($\nu<200$~MHz). Without calibration and/or filtering of this noise, current and future instruments may not be able to attain their theoretical sensitivity limit.  \\

Ionospheric propagation effects are direction dependent, and have traditionally been mitigated using self-calibration \citep{selfcal}. Self-calibration is very effective on individual sources observed with a narrow field of view (FOV). With a wide FOV of several to tens of degrees, there may not be enough signal to noise ratio, or worse yet, enough constraints to solve for phase errors in different directions within the relevant decorrelation time-scales. The residual direction-dependent errors will invariably manifest as scintillation noise in visibilities. Such propagation effects have long been identified as `challenges' to low frequency widefield observations. Yet, there has not been a concerted effort to evaluate the statistical properties of scintillation noise-- a primary aim of this paper.\\

Various aspects of radio wave propagation through turbulent plasma have been studied since the discovery of radio-star scintillation \citep{smith1950,hewish1952}. Earlier theoretical work concentrated mainly on understanding intensity scintillations \citep{mercier1962, salpeter1967} seen in total power measurements made with a zero baseline. With the advent of Very Long Baseline Interferometry (VLBI), investigations into the general case of visibility scintillation were carried out \citep{cronyn1972, goodman1989}. The above authors all assume a small FOV, and compute the statistics of scintillation for a single source that is unresolved, or partially resolved by the interferometer baseline-- a case that is not relevant for current and future arrays with wide FOVs of several to tens of degrees. Recently, \citet{koopmans2010} has taken into account a wide FOV, and a three-dimensional ionosphere to study the ensemble averaged visibilities that correspond to long exposures over which stable speckle-haloes or `seeing' develops around point-like radio sources. In this paper though, we are mainly concerned with second-order visibility statistics such as visibility variance, and the associated temporal, spectral, and spatial correlation properties of visibility scintillation for a wide FOV interferometer.\\

The rest of the paper is organised as follows. Section \ref{sec:basics} describes the basic properties of plasma turbulence, and its effect on the phase of electromagnetic waves. In Section \ref{sec:singlebas}, we compute the visibility statistics for a single baseline due to phase modulation by a turbulent plasma. In doing so, since we are generalising earlier results concerning scintillation of point-like sources to the case of an arbitrary sky intensity distribution, we have built on and/or expanded many of the algebraic deductions from the works of \citet{condona1986,coles1987,cronyn1972}. Where appropriate, we have included the deductions as applied to our case in the appendices for completeness. In Section \ref{sec:realsky}, we use the results of Section \ref{sec:singlebas} in conjunction with a realistic sky model, to make forecasts for visibility scintillation due to ionospheric propagation. We choose the ionospheric case, since it is the dominant source of scintillation in current low frequency radio telescopes. However, our notation is generic enough so as to be applicable also to interplanetary and interstellar scintillation. In Section \ref{sec:coherence}, we discuss the temporal, spatial, and spectral coherence of visibility scintillation-- properties that are important to the evaluation of time/frequency averaging and aperture synthesis effects. Finally, in Section \ref{sec:conclusions} we present our salient conclusions, and draw recommendations for future work.
%
%
%
%
\section{Basic properties}
\label{sec:basics}
A turbulent plasma introduces a time, frequency, and position dependent propagation phase on electromagnetic waves. These phase fluctuations are a direct consequence of density fluctuations in the plasma due to turbulence. Consequently, the propagation phase is expected to have certain statistical behaviour in time, frequency, and position. These statistical properties have been described in detail elsewhere (see \citet{wheelon1} and references therein), and we only summarise them here. We will make use of the widely used `thin screen' approximation \citep{ratcliffe1956}, wherein we assume the propagation phase in any given direction to be the integrated phase along that direction. This reduces the statistical description of plasma turbulence to an isotropic function in two dimensions.
\subsection{Frequency dependence}
The refractive index in a non-magnetised plasma is given by
\begin{equation}
\label{eqn:refindex}
\eta = \sqrt{1-\frac{\nu_{\rm p}^2}{\nu^2}} \approx 1-\frac{1}{2}\frac{\nu_{\rm p}^2}{\nu^2},
\end{equation}
where $\nu_{\rm p}$ is the electron plasma frequency, $\nu$ is the electromagnetic wave frequency, and the approximation holds for $\nu\gg\nu_p$. The plasma frequency itself is given by 
\begin{equation}
\label{eqn:plasmafreq}
\nu_{\rm p} = \frac{1}{2\upi} \sqrt{\frac{n_e e^2}{m_e \epsilon_0}},
\end{equation}
where $e$ and $m_e$ are the electron charge and mass respectively, and $\epsilon_0$ is the permittivity of free space. Typical ionospheric plasma frequency values are of the order of a few MHz. The phase shift due to wave propagation under the thin screen approximation is
\begin{equation}
\label{eqn:phase}
\phi_{\mathrm{tot}} = \int {\rm d} z \frac{2\upi \eta(z)}{\lambda},
\end{equation}
where $\lambda = c/\nu$ is the electromagnetic wavelength, $c$ is the speed of light in vacuum, and $z$ is the distance along the propagating ray. Using equation \ref{eqn:refindex}, we get
\begin{equation}
\phi_{\mathrm{tot}} = \int {\rm d} z \frac{2\upi  \nu}{c} -\frac{1}{2}\int {\rm d} z \frac{2\upi  \nu_{\rm p}^2}{c\nu}, 
\end{equation}
where the second term is the additional phase shift introduced due to the plasma: $\phi$ say, and the first term is a geometric delay that is usually absorbed into the interferometer measurement equation. It follows that the propagation phase $\phi$ is inversely proportional to the frequency $\nu$:
\begin{equation}
\label{eqn:freqscaling}
\phi(\nu) \propto \nu^{-1}\,\nu_{\rm p}^2.
\end{equation}
\subsection{Spatial properties} 
\begin{figure}
\centering
\includegraphics[width=0.9\linewidth]{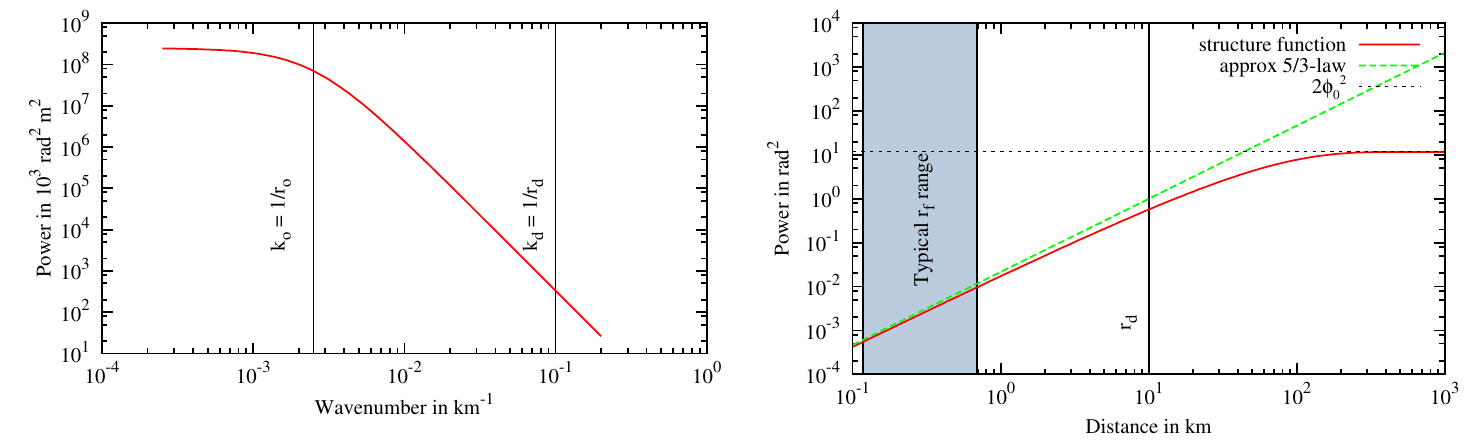}
\caption{Phase power spectrum (left panel) and the corresponding structure function (right panel) for typical values of ionospheric turbulence parameters: $r_{\rm o}=400$~km, $r_{\rm d}=10$~km, $\phi_0^2=5.87$~rad$^2$. The shaded region shows the range of Fresnel scale values for an ionospheric height of $300$~km at frequencies between $30$~MHz and $1$~GHz.  \label{fig:powspec}.}
\end{figure}

Spatial variations in plasma density $n_e$ may be modelled as a three-dimensional Gaussian random field with a power spectrum approximated by a $-11/3$ index power law corresponding to Kolmogorov-type turbulence\footnote{The statistics of ionospheric phase solutions in LOFAR data also attest this assumption (Mevius et al. priv. comm.).} \citep{ruffenach1972,singleton1974}. From equations \ref{eqn:plasmafreq} and \ref{eqn:freqscaling}, we have $\nu_{\rm p} \propto n_e^{1/2}$, and $\phi \propto \nu_{\rm p}^2$ respectively. It thus follows that $\phi \propto n_e$. Hence, the propagation phase is also a Gaussian random field with a power spectrum given by 
\begin{equation}
\ips{k} \propto k^{-11/3}\,\, k_{\rm o}<k<k_{\rm i},
\end{equation} 
where $k$ is the length of the spatial wavenumber vector $\mv{k}$, and $k_{\rm o}$ is the wavenumber corresponding to the outer scale or the energy injection scale, and $k_{\rm i}$ corresponds to the inner scale or energy dissipation scale. Note that we have assumed isotropy here for illustration, but we will keep the notation generic in the derivations so as to be applicable to an anisotropic power spectrum. We will assert the thin screen approximation by interpreting $k$ as the length of the spatial wavenumber vector in the two transverse dimensions, since $k_z=0$ essentially corresponds to the path integrated phase used in the thin screen approximation. For $k<k_{\rm o}$ the power spectrum is expected to be flat, and for $k>k_{\rm i}$ the power spectrum is expected to fall off rapidly to zero. For the ionospheric case, the inner scale is thought be to of the order of the ion gyroradius which is a few metres in length \citep{booker1979}. In the regime of interest to us, both the Fresnel scale which we defined later, and baseline lengths are significantly larger than the inner scale, and its effects may be safely ignored. In any case, the steep $-11/3$ index power law gives negligible power in turbulence on such small scales. The outer scale on the other hand, can be several tens to hundreds of kilometre. Such scales are typically within the projected field of view of current widefield telescopes on the ionosphere, and it is prudent to retain the effects of eddies on scales larger than the outer scale in widefield scintillation noise calculations. To make the computations analytically tractable, we will choose a form that has a graceful transition from the inertial $11/3$-law range for $k>k_{\rm o}$, and the flat range for $k<k_{\rm o}$\footnote{Our choice for the power spectrum is similar to the one made by \citet{vonkarman} in his study of fluid turbulence.}:
\begin{equation}
\label{eqn:vonkarman}
\ips{k} = \frac{5\phi_0^2 }{6\upi k_{\rm o}^2} \left[\left( \frac{k}{k_{\rm o}} \right)^2 +1 \right]^{-11/6},
\end{equation}
where we have normalised the spectrum to represent a two-dimensional Gaussian random field with variance $\phi_0^2$. We caution the reader that since there is no generally accepted theory of ionospheric plasma turbulence, neither the injection scale $k_{\rm o}$, nor the index ($\beta = 11/3$ here) are uniquely determined. We have chosen the $11/3$-law, since it corresponds to a well known Kolmogorov law, and since it falls within the range of $3<\beta<4$ suggested by measurements of ionospheric scintillation \citep{ruffenach1972}. The two-dimensional Fourier transform of equation \ref{eqn:vonkarman} gives the spatial autocorrelation function of the ionospheric phase:
\begin{equation}
\label{eqn:rhoiono}
\rho(r) = \frac{5}{3}\frac{(\upi k_{\rm o}r)^{5/6}}{\Gamma(11/6)} K_{\frac{5}{6}}(2\upi k_{\rm o}r),
\end{equation}
where $r$ is the spatial separation, $\Gamma(.)$ is the Gamma function, and $K_{\frac{5}{6}}(.)$ is the modified Bessel function of the second kind of order $\frac{5}{6}$. The autocorrelation function $\rho(.)$ has been normalised such that $\rho(0)=1$. For spatial separations significantly smaller than the outer scale ($rk_{\rm o}\ll 1$), we can use a small argument expansion of the Bessel function to get
\begin{equation}
\label{eqn:besseltaylor}
\rho(r) \approx \left[ 1-\frac{\Gamma(1/6)}{\Gamma(11/6)} (\upi k_{\rm o}r)^\frac{5}{3}\right].
\end{equation}
The spatial correlation is often described in terms of the structure function which is easier to measure in practice:
\begin{equation}
\label{eqn:structfunc}
\isf{\mv{r}} = \langle (\phi(\mv{r_0}+\mv{r}) - \phi(\mv{r_0}) )^2\rangle = 2\phi_0^2\left[ \rho(0) - \rho(r)\right].
\end{equation}
Using equation \ref{eqn:besseltaylor}, we can show that the structure function takes the usual form for Kolmogorov turbulence:
\begin{equation}
\label{eqn:structfuncrd}
\isf{r} \approx \left( \frac{r}{r_{\rm d}} \right)^{5/3},
\end{equation} 
where the approximation holds for $\upi r k_{\rm o} \ll 1$, and $\isf{r} \lesssim 2\langle \phi^2\rangle$, the latter being its asymptotic value, and $r_{\rm d}$ is the diffractive scale: the separation at which the phase structure function reaches unity. The diffractive scale is given by
\begin{equation}
\label{eqn:rd}
r_{\rm d} = \frac{1}{\upi k_{\rm o}} \left( \frac{\Gamma(11/6)}{2\Gamma(1/6)\phi_0^2}\right)^{3/5}.
\end{equation}
Finally, using the frequency scaling from equation \ref{eqn:freqscaling}, we can show that the diffractive scale varies with frequency as 
\begin{equation}
\label{eqn:diffscalefreq}
r_{\rm d}(\nu) \propto \nu^{6/5}.
\end{equation}
Typical values of the diffractive scale at $150$~MHz vary between $\sim 5$~km to $\sim 30$~km (Mevius et al. priv. comm.). Any two of the three variables $k_{\rm o}$, $\langle \phi^2 \rangle$, and $r_{\rm d}$ uniquely determine the power spectrum. Fig. \ref{fig:powspec} shows an isotropic power spectrum, and its structure function for typical ionospheric parameters specified at $150$~MHz: $r_{\rm o}=400$~km, $r_{\rm d}=10$~km, and $\phi_0^2=5.87$~rad$^2$. In the following sections, we will use a vector argument for the power spectrum and the structure function such that the results are also valid for anisotropic turbulence.\\
\subsection{Time dependence} 
The temporal variation in interferometric phase is usually dominated by the relative motion between the observer and the plasma irregularities, rather than an intrinsic evolution of the turbulence itself. For instance, ionospheric turbulence is expected to `ride along' a bulk wind at speeds of the order of $v=100$--$500$~km~hr$^{-1}$. This couples the temporal and spatial correlation properties of ionospheric phase, which we explore in Section \ref{sec:coherence}. Regardless, decorrelation of the ionospheric phase on a spatial scale $r$ implies a temporal decorrelation on a time-scale of 
\begin{equation}
\label{eqn:taud}
\tau_{\rm d} = r/v.
\end{equation}
As shown in Section \ref{subsec:tcorr}, the relevant spatial decorrelation scale is of the order of the baseline length with a minimum decorrelation scale equal to the Fresnel scale. For the case of ionospheric effects in current low frequency arrays, the above spatial scales vary from few hundred metres to several tens of kilometres. Hence, the relevant temporal decorrelation scales are of the order of few seconds to several minutes.
%
%
%
%
%
\section{Single baseline statistics}
\label{sec:singlebas}
\begin{figure}
\centering
\includegraphics[width=0.85\linewidth]{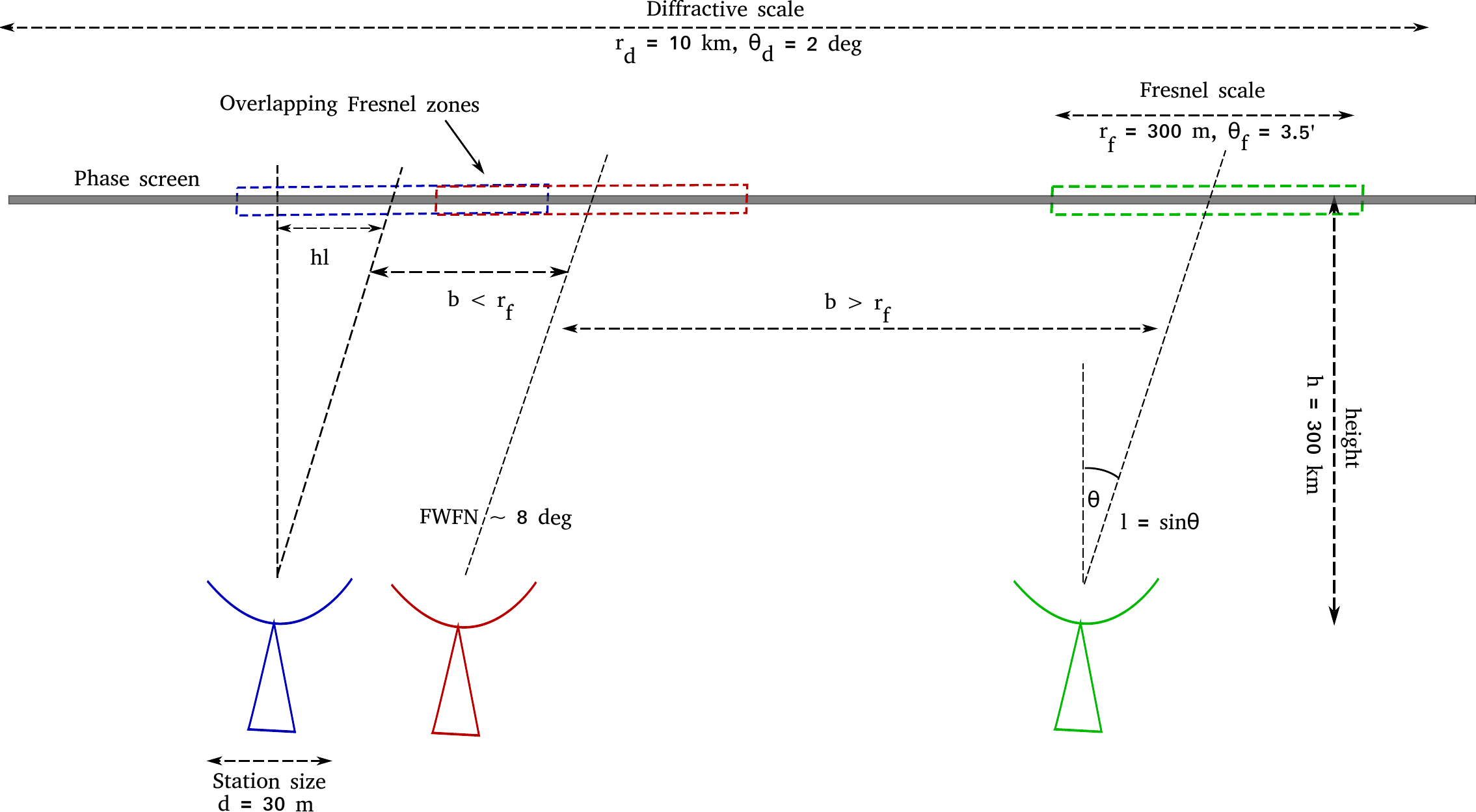}
\caption{A not-to-scale sketch showing the assumed geometry in this paper along with some length and angular scales that are relevant for our discussion. The numerical values are typical for the case of ionospheric propagation at $\nu=150$~MHz. \label{fig:geometry}}
\end{figure}

In this section, we derive the statistical properties of the interferometric visibility on a baseline formed by a given pair of antennas. We will assume that all antennas of the interferometer lie on a plane that is parallel to the diffraction screen, and denote all positions as vectors in two dimensions. The geometry is sketched in Fig. \ref{fig:geometry}.
The electric field on the observer's plane due to a unit flux source at position vector $\mv{l}$ is given by the Kirchhoff--Fresnel integral \citep{bornwolf} evaluated on the diffraction plane, which is the phase screen in our case:
\begin{equation}
\label{eqn:wffk}
E(\mv{r},\mv{l}) = \frac{1}{\ii \lambda h} \int {\rm d}^2\mv{x} \fe{(\mv{x}-\mv{r})^2}\,\, \me{-\ii 2\upi\mv{x}\bcdot \mv{l}/\lambda}\,\,  \me{\ii \phi(\mv{x})},
\end{equation}
where we have used the shorthand notation: $\mv{x}^2 = |\mv{x}|^2$. The second exponent accounts for the geometric delay in arrival times of the wavefront on different points on the diffraction plane, and the third exponent denotes the phase modulation of the wavefront as it crosses the phase screen\footnote{Taylor-expanding this exponential to $1^{\rm st}$ order in the weak-scattering regime gives the well-known Born approximation of the $1^{\rm st}$ order where $\phi(\mv{x})$ is the scattering amplitude.}. The first exponent which we will call the `Fresnel exponential', represents the effects of relative path-length differences between the `scatterers' on the diffraction screen at $\mv{x}$ and the observer at $\mv{r}$. Note that the relative scatterer--observer distance in equation \ref{eqn:wffk} is only accurate to quadratic order that corresponds to Fresnel diffraction. The higher order terms in the scatterer--observer distance become comparable to a wavelength if the FOV exceeds about $10$~deg. By completing the square in the first two exponents, we get
\begin{equation}
E(\mv{r},\mv{l}) = \frac{1}{\ii \lambda h} \me{-\ii 2\upi\mv{r}\bcdot \mv{l}/\lambda}\me{-\ii \upi h\mv{l}^2/\lambda} \int {\rm d}^2\mv{x} \fe{(\mv{x}-\mv{r}-h\mv{l})^2}\,\me{\ii \phi(\mv{x})}.
\end{equation}
Making a change of variable: $\mv{x}-\mv{r}-h\mv{l}  \rightarrow \mv{x} $, we get
\begin{equation}
\label{eqn:fk1}
E(\mv{r},\mv{l}) = \frac{1}{\ii \lambda h} \me{-\ii 2\upi\mv{r}\bcdot \mv{l}/\lambda} \me{-\ii \upi h\mv{l}^2/\lambda} \int {\rm d}^2\mv{x} \fe{\mv{x}^2} \me{\ii \phi(\mv{x}+\mv{r}+h\mv{l})},
\end{equation}
which is basically a convolution of the phase modulating function with the Fresnel exponential. The complex Fresnel exponential varies rapidly for $\mv{x}^2\gtrsim r^2_{\mathrm F}$ where $r_{\mathrm F}=\sqrt{\lambda h/(2\upi)}$ is called the Fresnel scale, and is depicted as dashed line rectangles in Fig. \ref{fig:geometry}. Consequently, most of the contribution to the integral comes from a small region of size $r_{\mathrm F}$ around the stationary phase point $\mv{x}=\mv{0}$. If the phase variation $\phi(\mv{x})$ on the diffraction screen is small ($\ll 1$~radian) over spatial scales of the size of $r_{\mathrm F}$, then the integral may be approximated by its value at the stationary phase point. This is often referred to as the pierce-point approximation, since we are reducing the electric field phase in a certain direction $\mv{l}$ to the ionospheric phase at $\mv{r}+h\mv{l}$, which is the point of intersection of a ray travelling from $\mv{r}$ in direction $\mv{l}$ with the scattering screen:
\begin{equation}
\label{eqn:pp1}
E_{\mathrm{pp}}(\mv{r},\mv{l}) = \me{-\ii 2\upi\mv{r}\bcdot \mv{l}/\lambda} \me{-\ii \upi h\mv{l}^2/\lambda} \me{\ii \phi(\mv{r}+h\mv{l})},
\end{equation}
where the subscript denotes the pierce-point approximation.\\

The visibility on a baseline $\mv{b}$ due to a source at $\mv{l}$ is defined as 
\begin{equation}
V(\mv{b},\mv{l}) \equiv E(\mv{r},\mv{l})E^{\ast}(\mv{r}+\mv{b},\mv{l}),
\end{equation}
where $(.)^{\ast}$ denotes complex conjugation. Since we assume the statistics of  the ionospheric phase to be spatially invariant, the visibility statistics are independent of the choice of $\mv{r}$, and we choose $\mv{r}$ to be the origin. Using the expression for the electric field from equations \ref{eqn:fk1} and \ref{eqn:pp1}, we can write the visibility for a unit flux-density source without and with the pierce-point approximation as
\begin{equation}
\label{eqn:fk}
V(\mv{b},\mv{l}) = \frac{\me{\ii 2\upi\mv{b}\bcdot \mv{l}/\lambda}}{\lambda^2 h^2} \int \int {\rm d}^2\mv{x_1} {\rm d}^2\mv{x_2} \fe{(\mv{x_1}^2-\mv{x_2}^2)}\,\me{\ii (\phi(\mv{x_1}+h\mv{l}) - \phi(\mv{x_2}+h\mv{l}+\mv{b}))}\,\,\, \textrm{and}
\end{equation}
\begin{equation}
V_{\mathrm{pp}}(\mv{b},\mv{l}) = \me{\ii 2\upi\mv{b}\bcdot \mv{l}/\lambda} \left[ \me{\ii (\phi(h\mv{l}) - \phi(h\mv{l}+\mv{b}))}\right], \,\,\, \mathrm{respectively.}
\end{equation}
Due to the convolution with the Fresnel exponential, the pierce-point approximation is accurate only when $b\gtrsim r_{\mathrm F}$ where the Fresnel zones for the two receiving antennas do not overlap (see Fig. \ref{fig:geometry}). In any case, the visibility from the entire sky can be written in terms of the point-source visibility as
\begin{equation}
V(\mv{b}) = \int \frac{{\rm d}^2\mv{l}}{\sqrt{1-\mv{l}^2}} I(\mv{l}) V(\mv{b},\mv{l}),
\end{equation}
where $I(\mv{l})$ is the apparent sky surface brightness as seen through the primary beam of the antennas comprising the interferometer elements. We are primarily interested in the statistical properties of $V(\mv{b})$ such as its expected value $\lr{V(\mv{b})}$, and variance $\sigma^2_V = \lr{|V(\mv{b})|^2} - \left| \lr{V(\mv{b})} \right|^2$. We want to compute these statistics as ensembles over different ionospheric phase screen realisations. The reader should not confuse these expectations with the expectations over the inherent randomness in emission from astrophysical sources, which has been made implicit in our notation. The expected value of the visibility is then given by
\begin{equation}
\lr{V(\mv{b})} =  \int \frac{{\rm d}^2\mv{l}}{\sqrt{1-\mv{l}^2}} I(\mv{l})\lr{V(\mv{b},\mv{l})}.
\end{equation}
The above expectation is analytically tractable and yields \citep[][see also Appendix \ref{sec:appa}]{bramley1955, ratcliffe1956}
\begin{equation}
\label{eqn:visexp}
\lr{V(\mv{b})} = \lr{V_{\mathrm{pp}}(\mv{b})} = \int \frac{{\rm d}^2\mv{l}}{\sqrt{1-\mv{l}^2}} I(\mv{l}) \me{\ii 2\upi\mv{b}\bcdot \mv{l}/\lambda}\me{-\frac{1}{2}\isf{\mv{b}}} = V(\mv{b})\me{-\frac{1}{2}\isf{\mv{b}}}.
\end{equation}
Hence, the expected visibility is equal to the visibility in the absence of the ionosphere, diminished by a factor that depends on the ionospheric phase structure function for a separation given by the baseline. Note that the above equation for the second moment of the electric field, is independent of the strength of scattering, and identical for both cases-- with and without the pierce-point approximation. As we will soon see, this similarity does not extend to higher moments of the electric field. \\

The visibility variance due to the entire sky is given by
\begin{equation}
\sigma^2\left[ V(\mv{b})\right] =  \int \frac{{\rm d}^2\mv{l}_{\rm a}}{\sqrt{1-\mv{l}_{\rm a}^2}} I(\mv{l}_{\rm a})\int \frac{{\rm d}^2\mv{l}_{\rm b}}{\sqrt{1-\mv{l}_{\rm b}^2}} I(\mv{l}_{\rm b}) \sigma^2\left[V(\mv{b},\mv{l}_{\rm a},\mv{l}_{\rm b}) \right].
\end{equation}
Analytically computing the two-source visibility variance ($\sigma^2\left[ V(\mv{b},\mv{l}_{\rm a},\mv{l}_{\rm b})\right]$) is tedious and not very enlightening. The interested reader may find the proof in Appendix \ref{sec:appb}, and we present the final expressions here:
\begin{equation}
\label{eqn:2srcvar}
\sigma^2\left[ V_{\mathrm{pp}}(\mv{b},\mv{l}_{\rm a},\mv{l}_{\rm b}) \right] = 4 \me{\ii 2\upi \mv{b}\bcdot \Delta \mv{l}/\lambda} \int {\rm d}^2\mv{q}\me{-\ii 2 \upi h \mv{q}\bcdot \Delta \mv{l}} \ips{\mv{q}} \sin^2\left( \upi \mv{q}\bcdot \mv{b}\right),\,\,\,\, \textrm{where }\Delta \mv{l} = \mv{l}_{\rm a}-\mv{l}_{\rm b},
\end{equation}
for the pierce-point approximation, and 
\begin{equation}
\sigma^2\left[ V(\mv{b},\mv{l}_{\rm a},\mv{l}_{\rm b}) \right] = 4 \me{\ii 2\upi \mv{b}\bcdot \Delta \mv{l}/\lambda} \int {\rm d}^2\mv{q}\me{-\ii 2\upi h\mv{q}\bcdot \Delta \mv{l}} \ips{\mv{q}} \sin^2 \left(-\upi\mv{q}\bcdot \mv{b} + \upi \lambda h\mv{q}^2 \right),
\end{equation}
for the full Kirchhoff--Fresnel integral. In deriving the above, we have assumed that the scattering is weak: the phase fluctuations within a Fresnel scale are small. The visibility variance is expressed as an integral of various wavemodes $\mv{q}$ in the phase power spectrum that are modulated by a sine-squared term which is a consequence of the Fresnel exponent. For this reason, this term is often called the Fresnel filter \citep{cronyn1972}. In Section \ref{subsec:intuition}, the Fourier domain representation will also be instrumental in developing a deeper intuitive understanding of Fresnel diffraction by a phase modulating screen. The pierce-point expression is a special case of the full Kirchhoff--Fresnel evaluation where the Fresnel scale in the Fresnel filter goes to zero-- a direct consequence of the stationary phase approximation. \\
 
\citet{cronyn1972} has derived an expression for visibility covariance between two redundant baselines that are spatially displaced by $\mv{d}$ and are looking at a single point-source. Whereas we are dealing with visibility covariance between two sources separated by $\Delta \mv{l}$, his expression is identical to our equation \ref{eqn:2srcvar} if we replace $h\Delta \mv{l}$ with $\mv{d}$. The similarity comes from the fact that both derivations are essentially evaluating the $4$-point correlation of ionospheric phase convolved with a Fresnel filter. In one case, the $4$ points are the pierce-points of the $4$ antennas forming the redundant baseline pair, each looking in some direction. In the other case, the pierce-points are those of the two antennas forming the baseline, looking in two different directions.\\

The visibility variance due to the entire sky can now be written as
\begin{equation}
\sigma^2 \left[ V(\mv{b})\right] = 4 \int \frac{{\rm d}^2\mv{l}_{\rm a}}{\sqrt{1-\mv{l}_{\rm a}^2}} I(\mv{l}_{\rm a})\int \frac{{\rm d}^2\mv{l}_{\rm b}}{\sqrt{1-\mv{l}_{\rm b}^2}} I(\mv{l}_{\rm b}) \me{\ii 2\upi \mv{b}\bcdot \Delta \mv{l}/\lambda} \int {\rm d}^2\mv{q}\me{-\ii 2\upi h \mv{q}\bcdot \Delta \mv{l}} \ips{\mv{q}} \sin^2 \left(-\upi\mv{q}\bcdot \mv{b} + \upi \lambda h\mv{q}^2 \right).
\end{equation}
Interchanging the order of integration, we get
\begin{equation}
\sigma^2 \left[ V(\mv{b})\right] = 4 \int {\rm d}^2\mv{q} \ips{\mv{q}} \sin^2 \left(-\upi\mv{q}\bcdot \mv{b} + \upi \lambda h\mv{q}^2 \right) \int \frac{{\rm d}^2\mv{l}_{\rm a}}{\sqrt{1-\mv{l}_{\rm a}^2}} I(\mv{l}_{\rm a})\int \frac{{\rm d}^2\mv{l}_{\rm b}}{\sqrt{1-\mv{l}_{\rm b}^2}} I(\mv{l}_{\rm b}) \me{\ii 2\upi (\mv{b}- \lambda h\mv{q}) \bcdot \Delta \mv{l}/\lambda }.
\end{equation}
The integrations with $\mv{l}_{\rm a}$ and $\mv{l}_{\rm b}$ yield the sky power spectrum\footnote{More precisely, the sky power spectrum in the absence of propagation effects.} computed at $\mv{b} - \lambda h\mv{q}$:
\begin{equation}
\label{eqn:skyps}
\int \frac{{\rm d}^2\mv{l}_{\rm a}}{\sqrt{1-\mv{l}_{\rm a}^2}} I(\mv{l}_{\rm a})\int \frac{{\rm d}^2\mv{l}_{\rm b}}{\sqrt{1-\mv{l}_{\rm b}^2}} I(\mv{l}_{\rm b}) \me{\ii 2\upi (\mv{b}- \lambda h\mv{q}) \bcdot \Delta \mv{l}/\lambda} = |V(\mv{b}-\lambda h\mv{q})|^2.
\end{equation}
Hence the visibility variance for the Kirchhoff--Fresnel evaluation is
\begin{equation}
\label{eqn:final}
\sigma^2 \left[V(\mv{b}) \right] = 4 \int {\rm d}^2\mv{q} \ips{\mv{q}} \sin^2 \left(-\upi\mv{q}\bcdot \mv{b} + \upi \lambda h\mv{q}^2 \right) |V(\mv{b}-\lambda h\mv{q})|^2,
\end{equation}
whereas the visibility variance for the pierce-point approximation is 
\begin{equation}
\label{eqn:final_pp}
\sigma^2 \left[V_{\mathrm{pp}}(\mv{b}) \right] = 4 \int {\rm d}^2\mv{q} \ips{\mv{q}} \sin^2 \left(\upi\mv{q}\bcdot \mv{b} \right) |V(\mv{b}-\lambda h\mv{q})|^2.
\end{equation}
We have thus related the visibility variance to the statistics of ionospheric turbulence-- via $\ips{\mv{q}}$, the scattering geometry-- via the Fresnel filter, and the sky power spectrum. We note here that equation \ref{eqn:final} is applicable to an arbitrary sky intensity power spectrum given by the $|V(\mv{b}-\lambda h\mv{q})|^2$ term. \citet[][equation 25]{cronyn1972} have derived an expression for visibility scintillation from a single source, where they make the assumption that source is unresolved by the interferometer baseline in the absence of propagation effects. Their equation for the scintillation variance is similar to our equation \ref{eqn:final}, but with the sky power spectrum replaced by $|V(\lambda h \mv{q})|^2$-- valid only with the unresolved source assumption. While this assumption is valid for scintillation of isolated compact sources such as pulsars and some quasars, it is not necessarily valid for the case of low frequency widefield interferometry due to the presence of sky emission on many spatial scales coming from a myriad of sources. \\

The pierce-point approximation leads to evident inconsistencies. For instance, when $\mv{b}=\lambda h \mv{q}$, the visibility variance receives a substantial contribution from the total power emission in the sky. In the Kirchhoff--Fresnel expression, however, the Fresnel filter vanishes for $\mv{b}=\lambda h\mv{q}$. However for $|\mv{b}|\gg r_{\mathrm F}$, the Fresnel filter term in equation \ref{eqn:final} reduces to the one in equation \ref{eqn:final_pp}. The pierce-point approximation works well for baselines far larger than the Fresnel scale, but gives erroneous results for baselines of the order of the Fresnel scale-- an important conclusion for current and future low frequency radio telescopes that have compact array configurations.
\subsection{Physical interpretation in one dimension}
\label{subsec:intuition}
\begin{figure}
\centering
\includegraphics[width=0.8\linewidth]{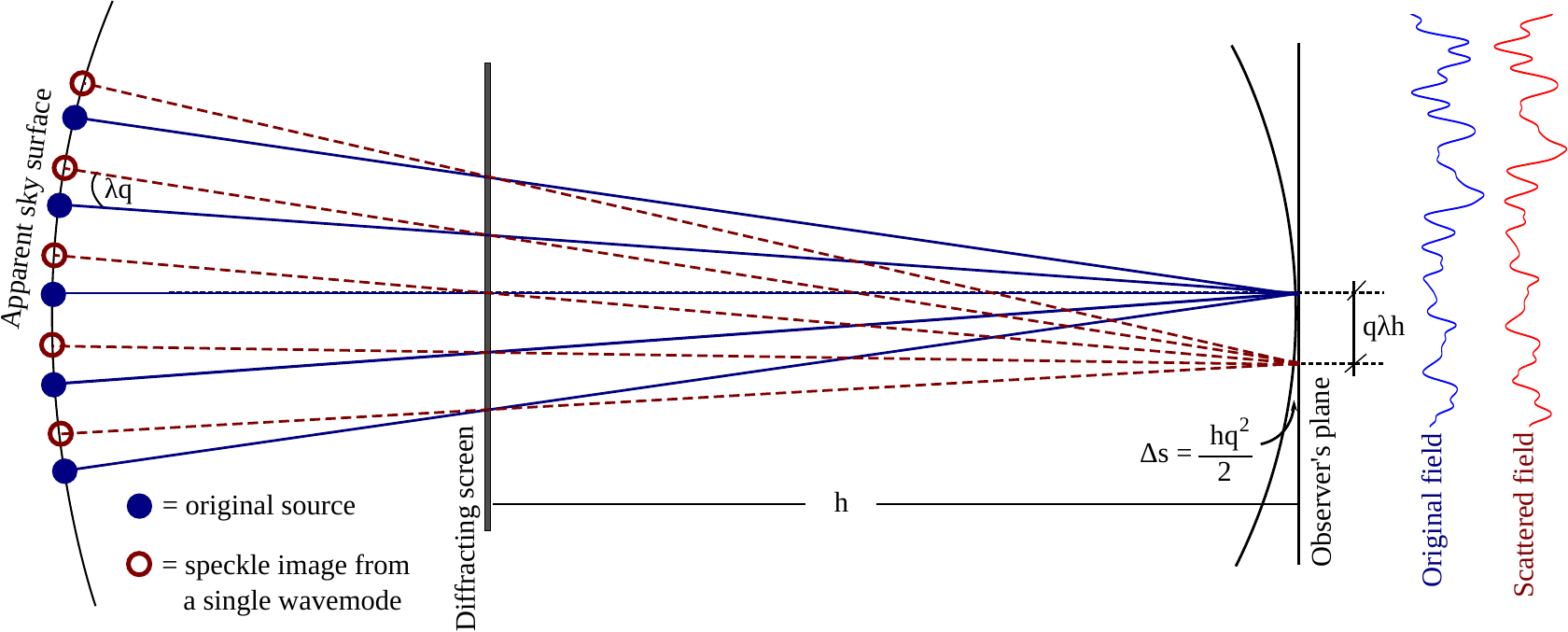}
\caption{Cartoon (not actual ray-tracing) depicting the physical interpretation of equation \ref{eqn:final}. A single ionospheric wavemode with spatial frequency $\mv{q}$ results in the displacement of the electric field on the observer's plane by an amount $\mv{q}\lambda h$. Equivalently, part of the flux in a source in the direction $\mv{l}$ is scattered into directions $\mv{l}+\mv{q}$ and $\mv{l}-\mv{q}$.\label{fig:interp}}
\end{figure}

We will now present some physical intuition behind equation \ref{eqn:final}. In doing so, our emphasis will be on the `meaning' or significance of the terms and not on the algebraic correctness. Hence, we will simply use a hypothetical one-dimensional sky and phase-screen. Equation \ref{eqn:final} is an integral on various Fourier modes-- with spatial frequency $q$-- of the modulating phase on the diffraction screen. The diffraction pattern on the observer's plane is a superposition of the Fresnel diffraction patterns due to each of these Fourier modes. The amplitudes of these Fourier modes are mutually independent: $\lr{\widetilde{\phi}(q_1)\widetilde{\phi}^{\ast}(q_2)} =0$ for $|q_1|\ne |q_2|$, and we can add the visibility variances due to individual Fourier modes as in equation \ref{eqn:final}. The electric field at position $r$ on the observer's plane $E(R)$ can be written in terms of the electric field on the diffraction plane $E_D(r)$ using the Kirchhoff--Fresnel integral:
\begin{equation}
E(R) = \frac{1}{\sqrt{\ii \lambda h }}\int {\rm d} r\, E_D(r) \fe{(r-R)^2} \me{\ii\phi(r)}.
\end{equation}
We will again make the weak-scattering approximation, and Taylor-expand the exponent containing the modulation phase $\phi(r)$ to write
\begin{equation}
E(R) = \frac{1}{\sqrt{\ii \lambda h }}\int {\rm d} r\, E_D(r) \fe{(r-R)^2} + \frac{\ii}{\sqrt{\ii \lambda h }}\int {\rm d} r\, E_D(r)\phi(r)\fe{(r-R)^2}.
\end{equation}
The first integral gives the electric field on the observer's plane in the absence of any scattering, say $E_{0}(R)$. The second term is the scattered field $E_s(R)$, and it is the interference between these two fields that we are interested in. $E_s(R)$ can be written by expressing $\phi(r)$ as a Fourier transform:
\begin{equation}
E(R) = E_{0}(R) + \frac{\ii}{\sqrt{\ii \lambda h}} \int {\rm d} q\, \widetilde{\phi}(q)\int {\rm d} r\, E_D(r) \fe{(r-R)^2} \me{\ii 2\upi qr}.
\end{equation}
Completing the square in the complex exponent, we get
\begin{equation}
E(R) = E_{0}(R) + \frac{\ii}{\sqrt{\ii \lambda h}} \int {\rm d} q\, \widetilde{\phi}(q) \me{\ii2\upi qR}\me{-\ii \upi \lambda h q^2}  \int {\rm d} r E_D(r) \fe{(r-R +\lambda hq)^2}.
\end{equation}
The second integral is equal to the incident field shifted by $\lambda h q$: $E_{0}(R-\lambda hq)$. Hence, we get
\begin{equation}
\label{eqn:final_interp}
E(R) = E_{0}(R) +\ii \int  {\rm d} q\,E_{0}(R-\lambda hq)\, \widetilde{\phi}(q) \me{\ii2\upi qR}\me{-\ii \upi \lambda h q^2}.
\end{equation}
The lateral shift of the scattered field on the observer plane is a direct consequence of weak phase modulation of the electric field on the diffraction plane by a `phase wave' with a spatial frequency of $q$. For instance, consider a plane wave travelling in direction $l$. Its geometric phase on the diffraction screen at position $r$ is $2\upi lr$. Phase modulation by a `phase wave' of spatial frequency $q$ adds an additional phase of $2\upi qr$. The aggregate phase is then $2\upi(l+q)r$-- that of a plane wave travelling in direction $l+q$. Hence, an incident wave from direction $l$ emerges from the diffraction plane travelling in direction $l+q$. This effect is depicted in Fig. \ref{fig:interp} where the sky is represented as a set of point-like sources denoted by filled blue circles on an imaginary `sky surface'. In the absence of the diffracting screen, the waves from these sources interfere to produce an instantaneous electric field on the observer's plane $E_{0}(R)$ depicted as a stochastic blue curve labelled `original field'. The diffracted waves, each being `deflected' by an angle $q$ form an interference pattern that is shifted on the observer's plane by an amount $\lambda qh$. This is depicted as the stochastic red curve labelled `scattered field' in Fig. \ref{fig:interp} . It is the interference between the direct incident field $E_{0}(R)$ and the stochastic\footnote{Stochastic here refers to the random nature of $\widetilde{\phi}(q)$.} scattered field $E_{0}(R-\lambda hq)$ that leads to most of the visibility scintillation noise. Due to a lateral shift of $\lambda hq$ between the interfering electric fields, visibility scintillation on a baseline $b$ is indeed sensitive to sky structures on baseline $b-\lambda hq$ as evidenced in equation \ref{eqn:final}. Finally, the additional geometric phase terms in equation \ref{eqn:final_interp} are a consequence of the additional path-length travelled by the deflected rays, which on including wavefront curvature effects, lead to the sine-squared term called the Fresnel filter in equation \ref{eqn:final}.  \\

We will demonstrate the above deductions more formally by considering a single wave mode: $\widetilde{\phi}(q) = \widetilde{\phi}(q_0)\delta(q-q_0) + \widetilde{\phi}^{\ast}(q_0)\delta(q+q_0)$, where $q_0>0$ and we have imposed conjugate symmetry to get a real phase field $\phi(r)$. The electric field on the observer's plane is then
\begin{equation}
E(R) = E_{0}(R) + \ii \widetilde{\phi}(q_{\rm o}) E_{0}(R-\lambda hq_0) \me{\ii 2\upi q_0R}\me{-\ii\upi\lambda hq_0^2} + \ii \widetilde{\phi}^{\ast}(q_0)E_{0}(R+\lambda hq_0) \me{-\ii 2\upi q_0R}\me{-\ii\upi\lambda hq_0^2}.
\end{equation}
The instantaneous visibility on baseline $b$ can be written as
\begin{equation}
V(b) = E(-b/2)E^{\ast}(b/2) = V_{0}(b) + 2\widetilde{\phi}^{\ast}(q_0)V_{0}(b-\lambda hq_0)\sin(-\upi q_0b+\upi\lambda hq_0^2) +2\widetilde{\phi}(q_0)V_{0}(b+\lambda hq_0)\sin(\upi q_0b+\upi\lambda hq_0^2),
\end{equation}
where we have disregarded the higher order terms in $\widetilde{\phi}(q_0)$ which can be shown to reduce to zero up to fourth-order in the visibility variance. The fourth-order terms are expected to be negligible for weak scattering. The first term-- $V_{0}(b)$ is the incident visibility in the absence of scattering, and the other terms are the result of interference between the incident and scattered fields. The variance of the visibility (over phase-screen realisations) may be computed by observing that $\lr{\left[\widetilde{\phi}(q_0)\right]^n}=\lr{\left[\widetilde{\phi}^{\ast}(q_0)\right]^n}=0$, for $n=1,2$ and $\lr{\widetilde{\phi}(q_0)\widetilde{\phi}^{\ast}(q_0)} = \ips{q_0}$:
\begin{equation}
\label{eqn:varq0}
\sigma^2\left[ V(b)\right] = \sigma^2\left[V_{0}(b) \right] + 4 \ips{q_0}\sin^2\left[-\upi q_0b+\upi\lambda hq_0^2\right]\left|V_{0}(b-\lambda hq_0) \right|^2 + 4 \ips{q_0}\sin^2\left[\upi q_0b+\upi\lambda hq_0^2\right]\left|V_{0}(b+\lambda hq_0) \right|^2
\end{equation}
where $q_0>0$. The term $\sigma^2\left[V_{0}(b) \right]$ is the visibility noise in the absence of scattering (sky noise + receiver noise), and the second term is the scintillation noise contribution to the visibility variance. Since the complex amplitude for different wavemodes $\widetilde{\phi}(q)$ are uncorrelated, we can express the total visibility variance as an integral of variances due to a individual wave modes as computed in equation \ref{eqn:varq0}:
\begin{equation}
\label{eqn:1deqnfinal}
\sigma^2\left[V(b) \right] = 4\int_{q=-\infty}^{q=+\infty} {\rm d} q\, \ips{q}\sin^2(-\upi qb+\upi\lambda hq^2)\left|V_{0}(b-\lambda hq) \right|^2\,\,\, (\textrm{scint. noise component})
\end{equation}
where we have extended the limits of integration to include negative values of $q$. Equation \ref{eqn:1deqnfinal} is a one-dimensional analogue of equation \ref{eqn:final}, but we derived it along with some physical intuition behind the nature of visibility scintillation.\\

An ionospheric wavemode of spatial frequency $q_0$ creates a coherent copy of the original sky but shifted by an angle $q_0$. The phase coherence between the original sky sources and their respective shifted copies leads to constructive and destructive interference on the observer's plane. The interference pattern varies due to turbulent fluctuations in the plasma screen, leading to visibility scintillation. The reader may note that this interference effect does not directly follow from application of the van Cittert--Zernike theorem often used in Fourier synthesis imaging, since it assumes that all sources are independent, and hence incoherent radiators. 
%
%
%
%
%
%
\section{Scintillation noise for a realistic sky model}
\label{sec:realsky}
As shown in equation \ref{eqn:final}, to compute the scintillation noise in visibilities, we need to know the sky power spectrum $|V(\mv{b})|^2$. The sky power spectrum obviously depends on the part of the sky being observed. However, we expected it to have certain average properties. On short baselines that are sensitive to large angular modes, the sky power spectrum is dominated by Galactic diffuse emission, and on longer baselines, the power spectrum is dominated by the contribution from a multitude of compact and point-like sources. Since the Fresnel filter vanishes for $\mv{b}\approx \lambda h\mv{q}$, we expect a sub-dominant contribution from the Galactic diffuse emission, and in this section, we numerically compute the scintillation noise due to point-like sources as a function of frequency and baseline length.\\

The sky power spectrum due to point-like sources can be written as
\begin{equation}
\label{eqn:vps}
|V(\mv{b})|^2 = \sum_{a=0}^{N-1} \sum_{b=0}^{N-1} S_{\rm a} S_{\rm b} \me{\ii 2 \upi \mv{b}\bcdot (\mv{l}_{\rm a}-\mv{l}_{\rm b})/\lambda},
\end{equation}
where we have assumed the sky to consist of $N$ sources, and the $i^{\rm th}$ source has a flux density $S_i$. Clearly, the sky power spectrum depends on the angular distribution of sources and their relative flux densities. For simplicity, we will assume that sources are distributed uniformly in the sky (no clustering). We will also assume that the average separation between sources $\mv{l}_{\rm a}-\mv{l}_{\rm b}$ is larger than the interferometer fringe spacing $\lambda /b$. In practice, this assumption implies that we count all sources within the interferometer fringe spacing as a single point-like source.  Under these assumptions, if there are many sources within each flux-density bin, then the complex exponential in equation \ref{eqn:vps} decorrelates in the summations unless $a=b$. For $a=b$, we get
\begin{equation}
\label{eqn:vps1}
|V(\mv{b})|^2 = \sum_{a=0}^{N-1} S_{\rm a}^2.
\end{equation}
Hence, scintillation noise due to many point-like sources is equal to the scintillation noise from a single point-like source with flux density 
\begin{equation}
S_{\mathrm{eff}}=\sqrt{\sum_{a=0}^{N-1} S_{\rm a}^2}.
\end{equation}
We note here that the above assumptions give a baseline independent power spectrum which is sometimes referred to as the `Poisson floor' in the sky power spectrum due to point-like sources. A few dominant sources in the field will lead to an interference pattern which may deviate significantly from this Poisson floor. However, bright sources present a large signal to noise ratio to calibrate the propagation phase within scintillation decorrelation frequency and time-scales, and hence, we do not compute their scintillation noise contributions, assuming that they have been largely calibrated and removed. It is the scintillation noise from the myriad of intermediate and low flux-density sources which may not be removed from direction-dependent calibration due to insufficient signal to noise ratio that we are concerned with. $S_{\mathrm{eff}}$ can be evaluated using the density function for sources within different flux-bins:
\begin{equation}
\label{eqn:dnds}
\frac{{\rm d}^2N(S_{\rm t})}{{\rm d} S_{\rm t}{\rm d} \Omega} = C\,\,  S_{\rm t}^{-\alpha}\nu^{-\beta} \,\,\,\,\,\, \textrm{Jy$^{-1}$sr$^{-1}$},  
\end{equation}
where $dN$ is the expected number of sources at frequency $\nu$ per unit solid angle whose flux densities lie within an interval $dS_{\rm t}$ about $S_{\rm t}$, $C$ is a normalising constant, and $\alpha$ and $\beta$ are typically positive, and depend on the flux-density range. Note that the above source count is defined for the true flux density, not the apparent flux density. The apparent flux density at position $\mv{l}$ on the sky is given by
\begin{equation}
S(\mv{l}) = S_{\rm t}(\mv{l})\beam,
\end{equation}
where $\beam$ is the primary beam factor at frequency $\nu$ in direction $\mv{l}$ for a primary aperture of diameter $d$. For our scintillation noise calculations, we are interested in the source counts for the primary-beam weighted sky $N(S)$ which is the number of sources in the visibly sky whose apparent flux densities lie in an interval $dS$ about $S$.
Integrating over the visible $2\upi$ solid angle, we can write
\begin{equation}
\frac{{\rm d} N(S)}{{\rm d} S}= \int\int_{2\upi}{\rm d} \Omega\frac{{\rm d}^2 N \left[S/\beam) \right]}{{\rm d} S_{\rm t} {\rm d} \Omega} \left|\frac{{\rm d} S_{\rm t}}{{\rm d} S}\right|,
\end{equation}
where, we have made a change of variables from $S_{\rm t}$ to $S$, with a simple scaling by the Jacobian. We can do this since the relationship between true and apparent flux is monotonic. Using the source counts from equation \ref{eqn:dnds}, we get
\begin{equation}
\frac{{\rm d} N(S)}{{\rm d} S} = CS^{-\alpha}\nu^{-\beta}\int\int_{2\upi} {\rm d} \Omega B^{\alpha-1}(d,\nu,\mv{l}).
\end{equation}
We can then define an effective beam as\footnote{For the typical value of $\alpha=2.5$, the effective beam $\beff$ is about $20$--$25$ per cent smaller than the area under the beam.} 
\begin{equation}
\label{eqn:beff}
\beff= \int\int_{2\upi} {\rm d} \Omega B^{\alpha-1}(d,\nu,\mv{l}),
\end{equation}
and write the number of sources in the visible sky with apparent flux densities between $S$ and $S+dS$ as
\begin{equation}
\frac{{\rm d} N(S)}{{\rm d} S} = C\beff S^{-\alpha}\nu^{-\beta}.
\end{equation}
We can now evaluate the relevant quantity-- $\seff = \sqrt{\sum S^2}$, using the source counts as
\begin{eqnarray}
\label{eqn:seff}
\seffsq &=& \int_{S_{\mathrm{min}}} ^{S_{\mathrm{max}}} {\rm d} S \frac{ {\rm d} N(S)}{ {\rm d} S} S^2 \nonumber \\ &=& \frac{C\beff \nu^{-\beta}}{3-\alpha} \left( S_{\mathrm{max}}^{3-\alpha} - S_{\mathrm{min}}^{3-\alpha} \right) \approx \frac{C\beff\nu^{-\beta}}{3-\alpha} S_{\mathrm{max}}^{3-\alpha},
\end{eqnarray}
where the approximation holds since $\alpha<3$, typically. This implies that most of the scintillation noise contribution comes from bright sources. It is then relevant to evaluate to what flux-density limit self-calibration is able to remove ionospheric effects on the brightest sources. This limit is array and field dependent, a detailed discussion of which is beyond the scope of this paper. We will however proceed by assuming that calibration completely removes scintillation noise on all sources that present a signal to noise ratio per visibility that is larger than some factor $\zeta$, where we compute the thermal noise for a visibility integration bandwidth and time of $\Delta\nu$ and $\Delta\tau$ respectively.
We attain a signal to noise ratio per visibility of $\zeta$ when
\begin{equation}
\label{eqn:smax_cal}
\smax = \zeta\frac{\sefd}{\sqrt{2\Delta\nu\Delta\tau}},
\end{equation}
where $\sefd$ is the system equivalent flux density. Finally, using this in equation \ref{eqn:seff}, we get the effective scintillating flux after removal of effects on bright sources as
\begin{equation}
\label{eqn:seff2_cal}
\seffsq = \frac{C\beff \nu^{-\beta}}{3-\alpha}\left( \frac{\zeta \sefd}{\sqrt{2\Delta\nu\Delta\tau}}\right)^{3-\alpha}.
\end{equation}
We will now compute numerical values of $\seff$ for a reference $d=30$~m aperture at $\nu=150$~MHz and provide scaling laws to compute $\seff$ for other values. Table \ref{tab:refval} gives the values of this reference parameter set. 
\begin{table}
\caption{Reference parameters for calculation of the effective scintillating flux\label{tab:refval}}
\centering
\begin{tabular}{lll}
\hline 
Parameter 		& Value 			& Comments\\ \hline\\
$d$ 			& $30$~m 		& Primary aperture diameter\\
$\nu$ 			& $150$~MHz 		& \\ 
SEFD 			& $1200$~Jy 		& For $T_{\rm sky}$ of 300~Kelvin (excludes receiver noise contribution)\\ 
$\alpha$		& 2.5				& Power-law index for differential source counts \citep[][Fig. 4a]{windhorst1985}\\
$\beta$			& 0.8				& Average low frequency radio source spectral index \citep[][Fig. 7]{vlssr}\\
$\zeta$			& 5					& Ensures reliable calibration solutions\\ 
$B_{\rm eff}$	& $0.0033$~sr		& Numerical integration of equation \ref{eqn:beff} \\ 
$\Delta\nu$	& $1$~MHz			& Frequency cadence for calibration\\
$\Delta\tau$	& $2$~sec			& Typical scintillation decorrelation scale for short baselines\\
$S_{\rm max}$ (with cal)	& $3$~Jy			& Using equation \ref{eqn:smax_cal}\\
$S_{\rm eff}$ (with cal)	& $5.86$~Jy			& Using equation \ref{eqn:seff2_cal}\\
$S_{\rm max}$ (without cal)	& $3.52$~Jy			& Using equation \ref{eqn:smax_nocal}\\
$S_{\rm eff}$ (without cal)	& $6.1$~Jy			& Using equation \ref{eqn:seff2_nocal}\\ \hline

\end{tabular}
\end{table}
As will be shown in Sec. \ref{sec:coherence}, ionospheric effects decorrelate on time-scales of a few seconds on baselines of the order of the Fresnel scale ($r_{\rm F}=100$s of metres). The thermal noise per visibility for a $2$~sec, $1$~MHz integration is about $0.6$~Jy. For $\zeta=5$, this gives $S_{\rm max}(30\textrm{ m},150\textrm{ MHz}) = 3$~Jy. We can now scale the values for $\smax$ by noting that $\sefd$ varies with frequency and primary aperture diameter as $\nu^{-2.5}d^{-2}$. Hence, the scaling law for $S_{\rm max}$ from equation \ref{eqn:smax_cal} is
\begin{equation}
\smax = 3 \left(\frac{d}{30\textrm{ m}} \right)^{-2} \left( \frac{\nu}{150\textrm{ MHz}}\right)^{-2.5}\,\,\,\, {\rm Jy}
\end{equation}
We need to now choose suitable values for $C$, $\alpha$, and $\beta$ to evaluate $\seff$. Around this flux range (few to several Jy at $150$~MHz), based on the $1.4$~GHz source counts of \citet[][Fig. 4a]{windhorst1985} we will choose (see also table \ref{tab:refval}).\\
\begin{equation}
\label{eqn:our_source_counts}
\frac{{\rm d} N(S)}{{\rm d} S}= 3\times 10^3\, \left(\frac{\beff}{1\textrm{ sr}}\right)\,\left(\frac{S}{1\textrm{ Jy}}\right)^{-2.5} \left( \frac{\nu}{150\textrm{ MHz}}\right)^{-0.8}\,\,\, {\rm Jy}^{-1}
\end{equation}
where $\beta=0.8$ is the average spectral index with which the radio flux density scales with frequency \citep{vlssr}. Using the above source counts in equation \ref{eqn:seff} gives $S_{\rm eff}(30\textrm{ m}, 150\textrm{ MHz}) = 5.86$~Jy. We can then scale the value of $\seff$ for other values of $d$ and $\nu$ by assuming that the effective beam $\beff$ scales with $d$ and $\nu$ with the same law with which the area under the beam scales with $d$ and $\nu$, which is $d^{-2}\nu^{-2}$. Numerical evaluation of beam areas shows that the error we make in the ratio is below a few percent. With this assumption, using equation \ref{eqn:seff2_cal}, the scaling law for $S_{\rm eff}$ can be written as
\begin{equation}
\seff \approx 5.86\,\, \left( \frac{d}{30\,{\rm m}} \right)^{-1.5}\,\left(\frac{\nu}{150\textrm{ MHz}}\right)^{-2.025}\,\,\, \textrm{Jy}.
\end{equation}
\\

\begin{figure}
\centering
\includegraphics[width=\linewidth]{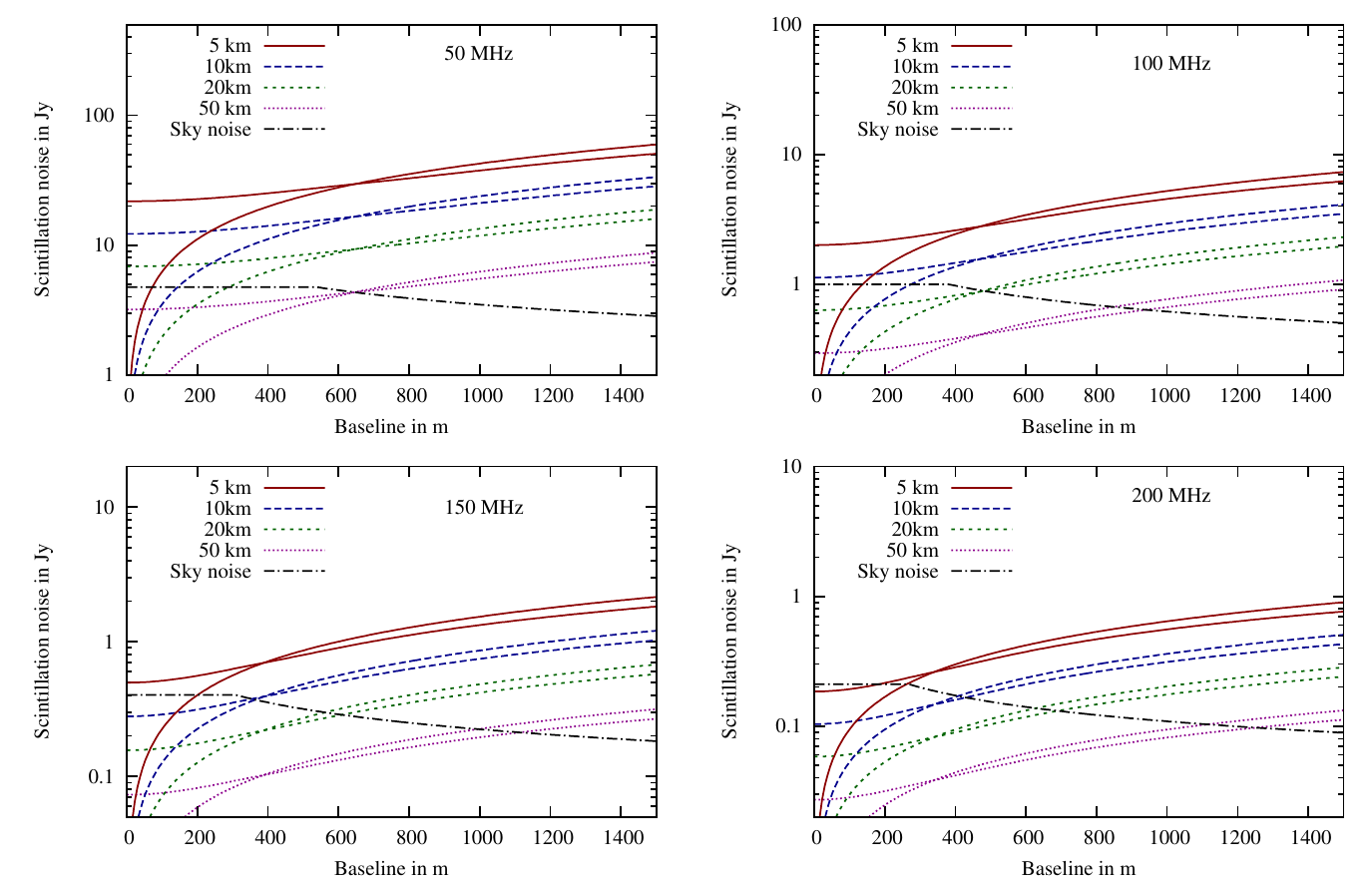}
\caption{Speckle noise rms (optimistic scenario) per snapshot visibility for different ionospheric diffractive scales specified at $150$~MHz, for a realistic source distribution and a primary aperture diameter of $30$~m. For each diffractive scale value, the curve that flattens at short baselines corresponds to the full Kirchhoff--Fresnel solution, while the curve that approaches zero at short baselines is computed using the pierce-point approximation. The different panels are for different frequencies ($50$, $100$, $150$, and $200$~MHz). Also shown for comparison (solid black) is the sky noise in visibilities assuming an integration over $1$~MHz in frequency, and the scintillation noise decorrelation time-scale in time. \label{fig:speckle_ps}}
\end{figure}

Fig. \ref{fig:speckle_ps} shows scintillation noise rms estimates as a function of baseline length for $S_{\mathrm{eff}}=5.86$~Jy (at $\nu=150$~MHz, $d=30$~m), and isotropic turbulence of the form given in equation \ref{eqn:vonkarman}. The four panels are for different frequencies between $50$ and $200$~MHz, and the different solid lines show the scintillation noise for a range of ionospheric diffractive scales (specified at $150$~MHz) typical to the LOFAR site (Mevius et al. priv. comm.) situated at mid-latitudes. The dashed lines show the scintillation noise computed using the pierce-point approximation, which as discussed before, gives inaccurate results at baselines $\lesssim r_{\rm F}$. Also shown in the figure are the thermal noise (sky noise only) for a $30$~m primary aperture, assuming an integration bandwidth of $1$~MHz, and integration time corresponding to the scintillation-noise decorrelation time-scale for each baseline (computed in Section \ref{subsec:tcorr}). Since $S_{\rm eff}(\nu)$ and the thermal noise do not scale with highly disparate indices ($-2.025$ and $2.5$ respectively), we expect the majority of spectral variation in thermal to scintillation noise ratio to be a result of increasing scattering strength with decreasing frequency.\\ 

The scintillation noise values in Fig. \ref{fig:speckle_ps} are computed assuming perfect removal of scintillation noise from all sources brighter than $S_{\rm max}(\nu,d=30) = 3(\nu/150\textrm{ MHz})^{-2.5}$~Jy using direction-dependent calibration. Since scintillation noise is dominated by the brighter sources in the field, the reader should interpret Fig. \ref{fig:speckle_ps} as an optimistic scenario.\\

It is also instructive to compute the effective scintillating flux in the absence of any calibration, or equivalently, if the calibration solutions are obtained with a temporal cadence that is significantly larger than the scintillation decorrelation time-scale. For such cases, we will choose $S_{\rm max}$ to be apparent flux-density threshold above which we expect to find, on an average, one source in the sky. The number of sources with apparent flux densities above $\smax$ is given by
\begin{equation}
N(S>S_{\rm max}) = \int_{S_{\rm max}}^{\infty} \frac{{\rm d} N(S)}{{\rm d} S} \approx \frac{C\nu^{-\beta}B_{\rm eff}(\nu)}{\alpha-1}S_{\rm max}^{1-\alpha }
\end{equation}
For $N(S>S_{\rm max}) = 1$, we get
\begin{equation}
\label{eqn:smax_nocal}
\smax = \left( \frac{\alpha-1}{C\nu^{-\beta}\beff}\right)^{1/(1-\alpha)}.
\end{equation}
For the source counts of equation \ref{eqn:our_source_counts}, we get $S_{\rm max}(30\textrm{ m},150\textrm{ MHz})=3.52$~Jy. We can write the scaling law for $S_{\rm max}$ as
\begin{equation}
S_{\rm max} = 3.52\left( \frac{\nu}{150\textrm{ MHz}}\right)^{-1.87} \left( \frac{d}{30\textrm{ m}}\right)^{-1.33}\,\,\textrm{ Jy}.
\end{equation}
Using equation \ref{eqn:seff}, The corresponding value for $S_{\rm eff}$ is then given by
\begin{equation}
\label{eqn:seff2_nocal}
S^2_{\rm eff} = \frac{(\alpha-1)^{(3-\alpha)/(1-\alpha)}}{3-\alpha} \left( CB_{\rm eff}(\nu)\nu^{-\beta}\right)^{2/(\alpha-1)},
\end{equation}
which yields $S_{\rm eff}(30\textrm{ m}, 150\textrm{ MHz}) = 6.1$~Jy, and the associated scaling law is
\begin{equation}
S_{\rm eff} = 6.1 \left( \frac{\nu}{150\textrm{ MHz}}\right)^{-1.87} \left(\frac{d}{30\textrm{ m}} \right)^{-1.33}\,\,\textrm{ Jy}.
\end{equation}
The effective scintillating flux in the absence of calibration is very close to that with calibration, attesting to the inefficacy of traditional self-calibration\footnote{By traditional we imply a channel by channel ($\Delta\nu\sim 1$~MHz) solution.} in mitigating scintillation noise. As shown in section \ref{subsec:freqcoh}, scintillation noise is a broadband phenomena in the weak-scintillation regime, and improved calibration algorithms that exploit the frequency coherence in scintillation noise are required to reduce scintillation noise by a significant amount. We also caution the reader here that the equations and arguments in this section give an ensemble value for $S_{\rm eff}(d,\nu)$. Since significant sample variance may exist in the actual number of bright sources in any field, a more representative value of $S_{\rm eff}$ for a particular field may be computed from an actual catalogue of sources in that field.\\

%
%
%
%
%
%
\section{Coherence properties of scintillation noise}
\label{sec:coherence}
So far, we have derived the statistical properties of visibility scintillation due to propagation though a turbulent plasma. These statistics must be interpreted as those for a quasi-monochromatic snapshot case, which refers to visibilities measured with an infinitesimal bandwidth and integration time. In reality, visibilities are always measured with certain spatial, temporal, and spectral averaging. Additionally, aperture synthesis results in averaging of visibilities on all the above dimensions. Accounting for these averaging effects requires knowledge of coherence properties of visibility scintillation in all three dimensions.
\subsection{Temporal coherence}
\label{subsec:tcorr}
\begin{figure}
\centering
\includegraphics[width=0.8\linewidth]{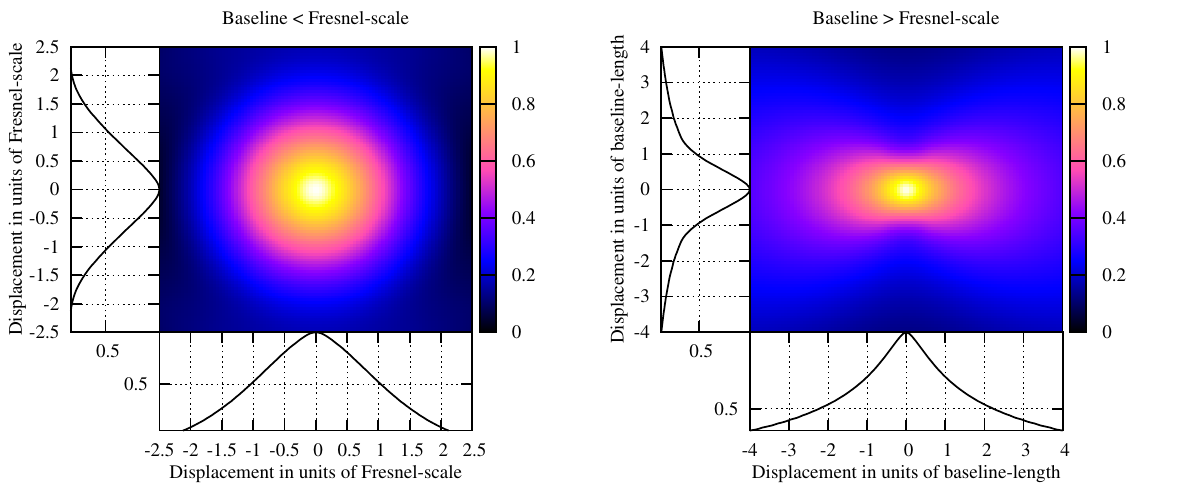}
\caption{Plot showing the correlation properties of scintillation noise from point-like sources as a function of displacement along (horizontal axis) and perpendicular (vertical axis) to the interferometer baseline. Displacement can be due to bulk motion of plasma-turbulence, or lateral shift of the baseline vector.  Left and right panels show the correlation when the interferometer-baseline is smaller than or larger than the Fresnel scale respectively.\label{fig:tcorr_2cases}}
\end{figure}

Temporal decorrelation of phase is expected to be mainly driven by the bulk motion of plasma turbulence relative to the observer, rather than the evolution of the turbulence itself. The visibility at time $t$ can be written as (making the time argument explicit):
\begin{equation}
V(\mv{b},\mv{l},t) = \frac{\me{\ii 2\upi\mv{b}\bcdot\mv{l}/\lambda}}{\lambda^2 h^2} \int \int {\rm d}^2\mv{x_1} {\rm d}^2\mv{x_2} \fe{(\mv{x_1}^2-\mv{x_2}^2)} \me{\ii (\phi(\mv{x_1}+h\mv{l}+\mv{v}t) - \phi(\mv{x_2}+h\mv{l}+\mv{b}+\mv{v}t))}
\end{equation}
where the vector $\mv{v}$ is the bulk wind velocity with which the `frozen' plasma irregularities move, and we have neglected the effects of varying baseline projection due to Earth rotation. The two-source visibility coherence on a temporal separation of $\tau$ is then
\begin{equation}
\sigma^2_\tau \left[V(\mv{b},\mv{l}_{\rm a},\mv{l}_{\rm b},\tau) \right] = \lr{V(\mv{b},\mv{l}_{\rm a},t=0)V^{\ast}(\mv{b},\mv{l}_{\rm b},t=\tau)}
\end{equation}
The derivation of the above temporal covariance follows the same steps as the ones in Appendix \ref{sec:appb} with $h\Delta \mv{l}$ replaced by $h\Delta \mv{l}+\mv{v}\tau$. Hence, we can write
\begin{equation}
\sigma^2_\tau \left[V(\mv{b},\mv{l}_{\rm a},\mv{l}_{\rm b},\tau) \right] = 4 \int {\rm d}^2\mv{q}\me{-\ii 2\upi (h\mv{q}\bcdot \Delta \mv{l}+\mv{q}\bcdot \mv{v}\tau)} \ips{\mv{q}} \sin^2 \left(-\upi\mv{q}\bcdot \mv{b} + \upi \lambda h\mv{q}^2 \right)
\end{equation}
The visibility variance due to the entire sky can now be written as (similar to equation \ref{eqn:final})
\begin{equation}
\label{eqn:tcorr}
\sigma^2_\tau \left[V(\mv{b},\tau) \right] = 4 \int {\rm d}^2\mv{q} \ips{\mv{q}} \sin^2 \left(-\upi\mv{q}\bcdot \mv{b} + \upi \lambda h\mv{q}^2 \right) |V(\mv{b}-\lambda h\mv{q})|^2 \me{-\ii 2 \upi \mv{q}\bcdot \mv{v}\tau}
\end{equation}
which is basically a Fourier transform relationship with $\mv{q}$ and $\mv{v}\tau$ as Fourier conjugates. This makes sense, since a lateral displacement of plasma wavemodes by an amount $v\tau$ decorrelates their aggregate phase over a `bandwidth' of $\Delta q = 1/(v\tau)$.  The temporal decorrelation characteristics for the point-source contribution to visibilities is given by replacing $|V(\mv{x})|^2$ in equation \ref{eqn:tcorr} by $S_{\mathrm{eff}}^2$. The resulting integration can be done numerically, and we show the results\footnote{$\sigma^2_\tau[V(\mv{b})]$ is in general complex for $|\mv{b}|\lesssim r_{\rm F}$, but the imaginary part is small compared to the real part. In Fig. \ref{fig:tcorr_2cases} we plot the absolute value of $\sigma^2_\tau[V(\mv{b})]$.} in Fig. \ref{fig:tcorr_2cases} for two limiting cases: (i) $|\mv{b}|\lesssim r_{\mathrm F}$ where the $\upi \lambda h \mv{q}^2$ term in the argument of the sine-squared function dominates, and (ii) $|\mv{b}|\gtrsim r_{\mathrm F}$ where the $\upi\mv{q}\mv{b}$ term dominates. In the second case, the Fourier transform can also be carried out analytically to yield
\begin{equation}
\sigma^2_\tau\left[V(\mv{b})\right] \approx S_{\mathrm{eff}}^2\phi_0^2\left[ 2\rho(\tau\mv{v})-\rho(\tau\mv{v}-\mv{b})-\rho(\tau\mv{v}+\mv{b})\right],\,\,\, |\mv{b}|\gtrsim r_{\mathrm F},
\end{equation}
where $\rho(.)$ is the spatial autocorrelation function of the ionospheric phase (see equation \ref{eqn:rhoiono}). From Fig. \ref{fig:tcorr_2cases}, we see that when $|\mv{b}|\lesssim r_{\mathrm F}$ (case 1), the correlation time  ($\tau_{\rm corr}=2r_{\mathrm F}/v$) is dictated by the time it takes the turbulence to cross the Fresnel scale, and for $|\mv{b}|\gtrsim r_{\mathrm F}$ (case 2) the correlation time ($\tau_{\rm corr}=2b/v$ or $4b/v$ depending on projection) is dictated by the time it takes the turbulence to cross the baseline-length. The latter is due to the fact that the visibility phase on baseline $|\mv{b}|$ is dominated by plasma wavemodes of size $\sim |\mv{b}|$ that decorrelate on length scales of the same order. But in the former case, the convolution with the Fresnel exponent sets a minimum decorrelation scale (spatially) that is of the order $r_{\mathrm F}$. For typical values of $\nu = 150$~MHz, $h = 300$~km, $v = 100$--$500$~km/hr for ionospheric scintillation parameters, the decorrelation time for $|\mv{b}|<r_{\mathrm F} (\approx 300\textrm{ m})$ varies between $4$ and $22$~seconds respectively, whereas for $|\mv{b}|=2$~km ($|\mv{b}|>r_{\mathrm F}$) the decorrelation time varies between $30$ and $150$~sec for plasma motion perpendicular to the baseline, and twice as much for plasma motion parallel to the baseline.
\subsection{Spatial coherence}
\label{subsec:spatcoh}
In practice, we average redundant, or near-redundant baselines, and hence we will concern ourselves with visibility coherence between baseline pairs that are identical (same length and orientation) but are displaced by a vector $\mv{s}$. It is straightforward to show that the coherence relationship is then identical to the one in equation \ref{eqn:tcorr} but with $\tau \mv{v}$ replaced by $\mv{s}$. This is because, laterally shifting the ionosphere by $\mv{s}$ is equivalent to shifting the baseline by the same amount. Hence, we arrive at the following conclusion. For visibility scintillation of the point-like sources, we again have two cases: (i) if $|\mv{b}|\lesssim r_{\mathrm F}$, then redundant baselines separated by more than the Fresnel scale ($r_{\mathrm F}$) experience incoherent visibility scintillation, and (ii) for $|\mv{b}|\gtrsim r_{\mathrm F}$, the separation between redundant baseline pairs must exceed the baseline length itself for the scintillation to decorrelate. Consequently, in highly compact arrays where all baselines lie within the Fresnel length $r_{\rm F}$, all near-redundant baselines experience coherence scintillation noise.
\subsection{Frequency coherence}
\label{subsec:freqcoh}
Analytically computing the visibility covariance between two frequencies is algebraically cumbersome, and we will restrict ourselves to heuristic arguments based on the terms in equation \ref{eqn:final}. Firstly, the overall magnitude of the effect varies as a function of frequency (via $\ips{\mv{q}}$) due to the frequency-scaling of the diffractive scale. Apart from this bulk effect, we expect decorrelation on smaller bandwidths due to geometric effects. Since the interferometer fringe-spacing scales with frequency, even in the absence of scattering, we expect frequency decorrelation in the visibility on wavelength scales of $\Delta \lambda_{\rm fringe} = d\lambda /b$: visibilities at wavelengths separated by more than $\Delta \lambda_{\rm fringe}$ are typically not averaged coherently. An additional geometric effect is imposed by the Fresnel filter (the sine-squared term). We can compute this by evaluating equation \ref{eqn:1deqnfinal} for visibility correlation at wavelengths $\lambda_1$ and $\lambda_2$:
\begin{equation}
\sigma^2\left[V(b,\lambda_1,\lambda_2) \right] = 4\int {\rm d} q\, \ips{q}\sin(-\upi qb+\upi\lambda_1 hq^2)\sin(-\upi qb+\upi\lambda_2 hq^2)\lr{V_{0}(b-\lambda_1 hq)V^*_{0}(b-\lambda_2 hq) },
\end{equation}
where we have assumed a sufficiently small separation between $\lambda_1$ and $\lambda_2$, such that variation in $\ips{\mv{q}}$ can be ignored. Using $\lambda_0=(\lambda_1+\lambda_2)/2$, and $\Delta\lambda=\lambda_1-\lambda_2$, we can write
\begin{equation}
\sigma^2\left[V(b,\lambda_1,\lambda_2) \right] = 4\int {\rm d} q\, \ips{q}\lr{V_{0}(b-\lambda_1 hq)V^*_{0}(b-\lambda_2 hq) } \left[\sin^2(-\upi qb+\upi\lambda_0 hq^2)-\sin^2(\upi\Delta\lambda hq^2/2)\right]
\end{equation}
which is the same as the visibility variance at $\lambda_0$, but with a modified Fresnel filter (sine-squared) term. The additional term in the new Fresnel filter-- $\sin^2(\upi\Delta\lambda hq^2/2)$ reaches appreciable values only for $\Delta\lambda \gtrsim 1/(2hq^2)$. Hence contribution from turbulence on spatial scales smaller than $1/q = \sqrt{2h\Delta\lambda}$ is suppressed in the visibility covariance, whereas contribution from larger scale fluctuations are mostly unaffected due to a change in wavelength. Due to the steep $-11/3$ law followed by $\ips{q}$, variance contribution from $\Delta\lambda \gtrsim 1/(2hq^2)$ is negligibly small for $\Delta \lambda \lesssim \lambda_0$, and we conclude that decorrelation in the Fresnel filter term is sub-dominant to fringe decorrelation. In the image domain, this can be thought of as the following: the frequency decorrelation in the observed speckle pattern is mostly due to a variation in the instantaneous\footnote{Instantaneous here must be interpreted as being within the typical decorrelation time-scale.} point-spread function (PSF) with frequency, rather than a variation in the intrinsic speckle pattern itself. Current low frequency arrays typically have low filling factors, and suffer significant snapshot PSF decorrelation with frequency. We expect this to be a dominant cause of scintillation decorrelation in the Fourier plane ($uv$-plane) over $\Delta\lambda \approx d\lambda/b$, or equivalently, $\Delta\nu/\nu \approx d/b$.
%
%
%
%
%
\section{Conclusions and future work}
\label{sec:conclusions}
Several new and upcoming radio telescopes operate at low radio frequencies ($\nu \lesssim 200$~MHz), and cater to a wide variety of science goals. The low frequencies and the accompanying wide fields-of-view require us to revisit plasma propagation effects that were earlier studied for the special case of observations of a single unresolved (or partially resolved) source at the phase-centre. We have done so in this paper, and have arrived at the following conclusions.
Propagation through a plasma (such as the ionosphere) imposes a frequency, time, and position dependent phase. The inherent randomness in plasma turbulence results in a stochastic visibility scintillation effect. We have derived expressions (equation \ref{eqn:final}) for the ensuing visibility variance for a wide field of view (several to tens of degrees) radio interferometer observing a sky with an arbitrary intensity distribution. Using these expressions, we show that for current low frequency arrays ($\nu \lesssim 200$~MHz) this source of uncertainty is typically comparable to, and in some regimes, larger than sky noise (Fig. \ref{fig:speckle_ps}).\\

The coherence time-scale for visibility scintillation of point-like sources is dictated by the time it takes for the turbulence to travel a distance $s=2b$ or $s=4b$ ($b$ is the baseline length) depending on whether the bulk velocity is perpendicular or parallel to the baseline. However, the coherence time cannot be smaller than the time it takes for the bulk motion to travel a distance of $s=2r_{\mathrm F}$, where $r_{\rm F}$ is the Fresnel scale. Coherence of visibility scintillation between redundant baseline pairs separated by $s$ is similar to temporal coherence on a time-scale of $\tau = s/v$. Due to their low filling factors, frequency decorrelation of visibility scintillation in current arrays is mostly cased by scaling of the snapshot point-spread function with frequency, rather than an evolution in the scintillation pattern itself.\\

Visibility scintillation effects are particularly relevant for experiments requiring high dynamic range measurements such as observations of the highly redshifted $21$-cm signal from the Cosmic Dawn and Reionization epochs. In this paper, we have made the first inroads into assessing the level of visibility scintillation in such experiments. The final uncertainty due to ionospheric propagation effects depends on the telescope geometry, and the extent to which calibration algorithms and other data processing operations can mitigate the above effects. We reserve a detailed discussion of these issues to a forthcoming paper.

\bibliographystyle{mn2e}
\bibliography{mybib.bib}
%
%
%
%
%
\appendix
\section{Single source visibility expectation}
\label{sec:appa}
Using equation \ref{eqn:visexp}, the single source visibility expectation is
\begin{equation}
\lr{V(\mv{b},\mv{l})} = \frac{\me{\ii 2\upi\mv{b}\bcdot \mv{l}/\lambda}}{\lambda^2 h^2} \int \int {\rm d}^2\mv{x_1} {\rm d}^2\mv{x_2} \fe{(\mv{x_1}^2-\mv{x_2}^2)}.\lr{ \me{\ii (\phi(\mv{x_1}+h\mv{l}) - \phi(\mv{x_2}+h\mv{l}+\mv{b}))}},
\end{equation}
an expression for which was provided by \citet{bramley1955, ratcliffe1956}. We include the proof here to introduce some algebraic concepts that will be used later. To compute the expectation on ionospheric phases, we will use the following theorem from \citet{mercier1962}: If $a_k$ are scalars, and $\phi_k$ are Gaussian random variables, then
\begin{equation}
\label{eqn:lemma1}
\lr{\me{\ii \sum_k a_k\phi_k}} = \me{-\frac{1}{2}\sum_k\sum_m a_ka_m\lr{\phi_k\phi_m}}
\end{equation}
The visibility expectation is then
\begin{equation}
\lr{V(\mv{b},\mv{l})} = \frac{\me{\ii 2\upi\mv{b}\bcdot \mv{l}/\lambda}}{\lambda^2 h^2} \int \int {\rm d}^2\mv{x_1} {\rm d}^2\mv{x_2} \fe{(\mv{x_1}^2-\mv{x_2}^2)}\me{-\phi_0^2(1-\rho(\mv{x_1}-\mv{x_2}-\mv{b}))}.
\end{equation}
Making the change of integration variables from $\mv{x_1},\mv{x_2}$ to $\mv{u},\mv{v}$ where $\mv{u} = (\mv{x_1}+\mv{x_2})/\sqrt{2}$ and $\mv{v} = (\mv{x_1}-\mv{x_2})/\sqrt{2}$, we get
\begin{equation}
\lr{V(\mv{b},\mv{l})} = \frac{\me{\ii 2\upi\mv{b}\bcdot \mv{l}/\lambda}}{\lambda^2 h^2} \int \int {\rm d}^2\mv{u} {\rm d}^2\mv{v} \fe{\mv{u}\bcdot \mv{v}}\me{-\phi_0^2 (1-\rho(\mv{v}\sqrt{2}-\mv{b}))}.
\end{equation}
The integration with respect to $\mv{u}$ is straightforward and yields, $\lambda^2 h^2\delta(\mv{v})$, where $\delta(.)$ is the two-dimensional Dirac-delta function. The integration with respect to $\mv{v}$ returns the integrand at $\mv{v}=\mv{0}$:
\begin{equation}
\lr{V(\mv{b},\mv{l})}  = \me{\ii 2\upi\mv{b}\bcdot \mv{l}/\lambda} \me{-\phi_0^2(1-\rho(\mv{b}))}.
\end{equation}
The result can be written in terms of the structure function $\isf{\mv{b}} = 2\phi_0^2(1-\rho(\mv{b}))$ as
\begin{equation}
\label{eqn:1srcexp}
\lr{V(\mv{b},\mv{l}) } = \me{\ii 2\upi\mv{b}\bcdot \mv{l}/\lambda} \me{-\frac{1}{2}\isf{\mv{b}}}.
\end{equation}
%
%
%
%
%
%
\section{Two-source visibility covariance}
\label{sec:appb}
\begin{figure}
\centering
\includegraphics[width=.6\linewidth]{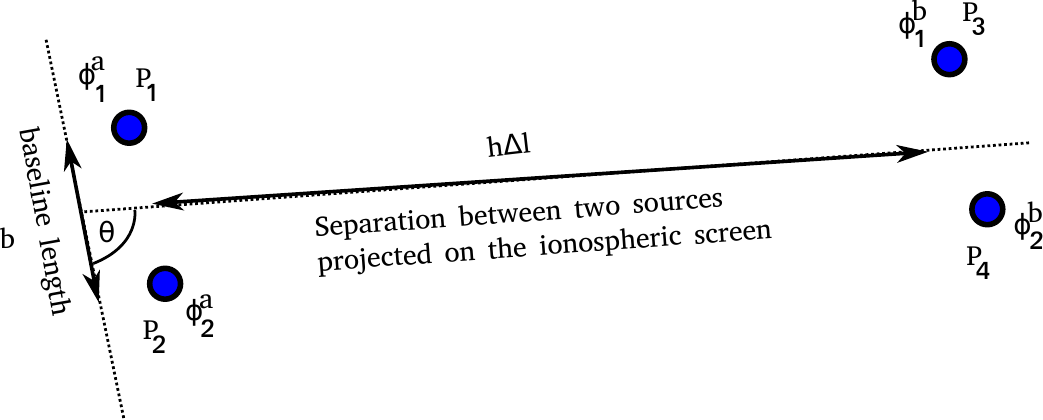}
\caption{Sketch comparing the baseline length to the projected separation (on the ionospheric screen) of the baseline for two sources \label{fig:piercepoints}}
\end{figure}

We define the two-source visibility covariance as
\begin{equation}
\label{eqn:2terms}
\sigma^2\left[ V_{\mathrm{pp}}(\mv{b},\mv{l}_{\rm a},\mv{l}_{\rm b}) \right] = \lr{V(\mv{b},\mv{l}_{\rm a})V^{\ast}(\mv{b},\mv{l}_{\rm b})} - \lr{V(\mv{b},{l_{\rm a}})}\lr{V(\mv{b},{l_{\rm b}})}^{\ast}.
\end{equation}
The first term is basically the mutual coherence between visibilities on the same baseline due to two sources in the sky:
\begin{eqnarray}
\lr{V(\mv{b},{l_{\rm a}})V^{\ast}(\mv{b},\mv{l}_{\rm b})} &=& \frac{\me{\ii 2 \upi \mv{b}\bcdot \Delta\mv{l}}}{\lambda^4h^4} \int\int\int\int {\rm d}^2\mv{x_1}{\rm d}^2\mv{x_2}{\rm d}^2\mv{x_3}{\rm d}^2\mv{x_4} \fe{(\mv{x_1}^2-\mv{x_2}^2-\mv{x_3}^2+\mv{x_4}^2)} \nonumber \\
& &  \lr{\me{\ii \left(\phi(\mv{x_1+h\mv{l}_{\rm a}})-\phi(\mv{x_2}+h\mv{l}_{\rm a}+\mv{b})-\phi(\mv{x_3}+h\mv{l}_{\rm b})+\phi(\mv{x_4}+h\mv{l}_{\rm b}+\mv{b}) \right)}}.
\end{eqnarray}
The expectation in the above equation is the $4$-point phase coherence on the ionospheric screen. Fig. \ref{fig:piercepoints} depicts the geometry of the $4$-points that correspond to the `pierce-points' on the ionospheric plane of the rays that go from the two antennas towards the two sources. The expectation in the above equation depends on the phase structure on all $16$ pairs that can be drawn from $4$ pierce-points, and can be written using equation \ref{eqn:lemma1} as
\begin{eqnarray}
\label{eqn:monster}
\lr{V(\mv{b},{l_{\rm a}})V^{\ast}(\mv{b},\mv{l}_{\rm b})} &=& \frac{\me{\ii 2 \upi \mv{b}\bcdot \Delta\mv{l}}}{\lambda^4h^4} \int\int\int\int {\rm d}^2\mv{x_1}{\rm d}^2\mv{x_2}{\rm d}^2\mv{x_3}{\rm d}^2\mv{x_4} \fe{(\mv{x_1}^2-\mv{x_2}^2-\mv{x_3}^2+\mv{x_4}^2)} \nonumber \\
& &  \left(\me{-\frac{\phi_0^2 \left( \psi\right)}{2}} \right), \\
\end{eqnarray}
where $\psi$ is given by
\begin{equation}
\psi = 4 - 2\left(\rho(\mv{x_{12}}+\mv{b})+\rho(\mv{x_{13}}+h\Delta\mv{l}) -\rho(\mv{x_{14}}+h\Delta\mv{l}-\mv{b}) -\rho(\mv{x_{23}}+h\Delta\mv{l}+\mv{b}) +\rho(\mv{x_{24}}+h\Delta\mv{l}) + \rho(\mv{x_{34}}-\mv{b}) \right),
\end{equation}
where we have used the shorthand notation $\mv{x_{ij}=\mv{x_i}-\mv{x}_j}$. The integrations may not be carried out analytically. In the weak-scattering regime, we may proceed by Taylor-expanding the exponent about $0$ as
\begin{equation}
\me{-\frac{\phi_0^2\psi}{2}} \approx 1-\frac{\phi_0^2\psi}{2}.
\end{equation}
Now that the exponent has been linearised, equation \ref{eqn:monster} reduces to a sum of integrals, with each integral being a Fresnel integral of a two-point correlation function $\rho(.)$. All but two of the integrals can be evaluated using a procedure similar to the one in Appendix \ref{sec:appa}, and we get
\begin{equation}
\lr{V(\mv{b},{l_{\rm a}})V^{\ast}(\mv{b},\mv{l}_{\rm b})} =\me{\ii 2 \upi \mv{b}\bcdot \Delta\mv{l}} \left[1-2\phi_0^2 \left(1-\rho(\mv{b})\right) + \phi_0^2\left(2\rho(h\Delta\mv{l}) - T_1 - T_2\right) \right],
\end{equation} 
where $T_1$ and $T_2$ have $\rho(\Delta\mv{x_{23}}+h\Delta\mv{l}+\mv{b})$ and $\rho(\Delta\mv{x_{14}}+h\Delta\mv{l}-\mv{b})$ as the integrands respectively. $T_1$ can be further reduced as follows.
\begin{equation}
T_1 =  \left[ \frac{1}{\lambda^2 h^2}\int\int {\rm d}^2\mv{x_1}{\rm d}^2\mv{x_4}\fe{(\mv{x_1}^2+\mv{x_4}^2)}\right].\left[\frac{1}{\lambda^2 h^2}\int\int {\rm d}^2\mv{x_2}{\rm d}^2\mv{x_3} \fe{(-\mv{x_2}^2-\mv{x_3}^2)} \rho(\Delta\mv{x_{23}}+h\Delta\mv{l}+\mv{b}) \right].
\end{equation}
The integrals with respect to $\mv{x_1}$ and $\mv{x_4}$ are both Fresnel integrals in the absence of any phase modulation, and each of them reduces to $\ii$, and their product is $-1$. To compute the integrals with respect to $\mv{x_2}$ and $\mv{x_3}$, we make the change of variables: $\mv{u} = (\mv{x_2}-\mv{x_3})/\sqrt{2}$, $\mv{v} = (\mv{x_2}+\mv{x_3})/\sqrt{2}$ to get
\begin{equation}
T_1 =  -\frac{1}{\lambda^2 h^2}\int\int {\rm d}^2\mv{u}{\rm d}^2\mv{v} \fe{(-\mv{u}^2-\mv{v}^2)} \rho(\sqrt{2}\mv{u}+h\Delta\mv{l}+\mv{b}).
\end{equation}
The integration with respect to $\mv{v}$ is again a Fresnel integral with no phase modulations and reduces to $-\ii$. Hence, we get 
\begin{equation}
\label{eqn:theend}
T_1 = \frac{\ii}{\lambda h}\int {\rm d}^2\mv{u} \fe{(-\mv{u}^2)} \rho(\sqrt{2}\mv{u}+h\Delta\mv{l}+\mv{b}).
\end{equation}
We are unable to reduce the integral analytically. However, equation \ref{eqn:theend} is a convolution between two functions at lag $\mv{b}+h\Delta \mv{l}$, and using the convolution theorem, we can write
\begin{equation}
\label{eqn:t1final}
T_1 = \frac{1}{\phi_0^2}\int {\rm d}^2\mv{q}\me{\ii 2 \upi \mv{q}\bcdot (\mv{b}+h\Delta \mv{l})} \ips{\mv{q}} \me{\ii 2 \upi \lambda h \mv{q}^2},
\end{equation}
where $\mv{q}$ and $h\Delta \mv{l}$ form a Fourier conjugate pair, $\ips{\mv{q}}$ is the Fourier transform of $\phi_0^2\rho(\mv{u})$, and $\me{\ii 2 \upi \lambda h \mv{q}^2}$ is the Fourier transform of $\ii/(\lambda h)\fe{(-\mv{u}^2)}$. Using a similar procedure, $T_2$ can be reduced to
\begin{equation}
T_2 = \frac{1}{\phi_0^2}\int {\rm d}^2\mv{q}\me{-\ii 2 \upi \mv{q}\bcdot (\mv{b}-h\Delta \mv{l})} \ips{\mv{q}} \me{-\ii 2 \upi \lambda h \mv{q}^2}.
\end{equation}
Hence $T_1+T_2$ is given by
\begin{equation}
T_1+T_2 = \frac{1}{\phi_0^2}\int {\rm d}^2\mv{q}\ips{\mv{q}}\me{\ii 2 \upi  h \mv{q}\bcdot \Delta \mv{l}} 2 \cos\left(2\upi \mv{q}\bcdot \mv{b} + 2\upi \lambda h \mv{q}^2\right).
\end{equation}
Collecting all terms, we get
\begin{equation}
\lr{V(\mv{b},{l_{\rm a}})V^{\ast}(\mv{b},\mv{l}_{\rm b})} = \me{\ii 2 \upi \mv{b}\bcdot \Delta\mv{l}} \left[1-2\phi_0^2\left(1-\rho(\mv{b})-\rho(h\Delta\mv{l})  + \int {\rm d}^2\mv{q} \ips{\mv{q}} \me{-\ii \upi  h \mv{q}\bcdot \Delta \mv{l}} \cos  \left(-2 \upi \mv{q} \mv{b}+2\upi\lambda h\mv{q}^2\right)\right) \right],
\end{equation}
where we have made the substitutions $\mv{q}\rightarrow -\mv{q}$ to preserve the sign convention in the Fourier transform with respect to $\mv{q}$. Writing $\phi_0^2\rho(h\Delta \mv{l})$ in terms of its Fourier transform, taking in into the integral, and using the trigonometric half-angle formula, we get
\begin{equation}
\label{eqn:covar}
\lr{V(\mv{b},{l_{\rm a}})V^{\ast}(\mv{b},\mv{l}_{\rm b})} = \me{\ii 2 \upi \mv{b}\bcdot \Delta\mv{l}} \left[1-2\phi_0^2+2\phi_0^2\rho(\mv{b}) + 4\int {\rm d}^2\mv{q}\me{-\ii 2\upi  h \mv{q}\bcdot \Delta \mv{l}} \ips{\mv{q}} \sin^2 \left(-\upi\mv{q}\bcdot \mv{b} + \upi \lambda h\mv{q}^2 \right) \right]
\end{equation}
The second term in equation \ref{eqn:2terms} can be evaluated using equation \ref{eqn:1srcexp} as
\begin{equation}
\lr{V(\mv{b},{l_{\rm a}})}\lr{V(\mv{b},{l_{\rm a}})}^{\ast} = \me{\ii 2\upi\mv{b}.\Delta \mv{l}/\lambda} \me{-\isf{\mv{b}}} = \me{\ii 2\upi\mv{b}.\Delta \mv{l}/\lambda} \me{-2\phi_0^2(1-\rho(\mv{b}))} 
\end{equation}
We may Taylor-expand the exponent in the weak-scattering limit to get
\begin{equation}
\label{eqn:mean}
\lr{V(\mv{b},{l_{\rm a}})}\lr{V(\mv{b},{l_{\rm a}})}^{\ast} = \me{\ii 2\upi\mv{b}.\Delta \mv{l}/\lambda} \left[ 1-2\phi_0^2+2\phi_0^2\rho(\mv{b})\right].
\end{equation}
Substituting equations \ref{eqn:covar} and \ref{eqn:mean}, in equation \ref{eqn:2terms}, we get the expression for the two-source visibility covariance:
\begin{equation}
\sigma^2\left[ V(\mv{b},\mv{l}_{\rm a},\mv{l}_{\rm b}) \right] = 4\me{\ii 2 \upi \mv{b}\bcdot \Delta\mv{l}}\int {\rm d}^2\mv{q}\me{-\ii 2\upi  h \mv{q}\bcdot \Delta \mv{l}} \ips{\mv{q}} \sin^2 \left(-\upi\mv{q}\bcdot \mv{b} + \upi \lambda h\mv{q}^2 \right)
\end{equation}
The two-source visibility covariance for the pierce-point approximation may be computed by discounting the Fresnel integrations in equation \ref{eqn:monster}, or in other words, by extracting the value of the integral at $\mv{x_1}=\mv{x_2}=\mv{x_3}=\mv{x_4}=\mv{0}$. The computations are straightforward, and yield
\begin{equation}
\sigma^2\left[V_{\mathrm{pp}}(\mv{b},\mv{l}_{\rm a},\mv{l}_{\rm b}) \right] = 4\me{\ii 2 \upi \mv{b}\bcdot \Delta\mv{l}}\int {\rm d}^2\mv{q}\me{-\ii 2\upi  h \mv{q}\bcdot \Delta \mv{l}} \ips{\mv{q}} \sin^2 \left(\upi\mv{q}\bcdot \mv{b}\right).
\end{equation}
%
%
%
%
%
%
\end{document}